\documentclass[
amsmath,
amssymb,
aps, 
pra,
twocolumn, 
groupedaddress,
nobalancelastpage,
floatfix]
{revtex4-2}
\usepackage{mathtools}
\usepackage[colorlinks]{hyperref}
\usepackage{lipsum}
\usepackage{bbold}
\usepackage{bm}
\usepackage{dsfont}
\usepackage{enumerate}
\usepackage{graphicx}
\usepackage{float}
\usepackage {xcolor}
\usepackage{cleveref}
\usepackage{amsmath}

\usepackage{lipsum}

\renewcommand{\vec}[1]{\bm{#1}}	
\newcommand{\vo}[1]{\hat{\vec{#1}}}

\DeclarePairedDelimiterX{\mean}[1]{\langle}{\rangle}{
	{#1}
}
\DeclarePairedDelimiterX{\abs}[1]{\lvert}{\rvert}{
	{#1}
}
\DeclarePairedDelimiterX{\norm}[1]{\lVert}{\rVert}{
	{#1}
}
\DeclarePairedDelimiterX{\bra}[1]{\langle}{\rvert}{#1}

\DeclarePairedDelimiterX{\ket}[1]{\lvert}{\rangle}{#1}

\DeclarePairedDelimiterX{\mel}[3]{\langle}{\rangle}{
	{#1}\delimsize\vert {#2}\delimsize\vert {#3}
}
\DeclarePairedDelimiterX{\inner}[2]{\langle}{\rangle}{
	{#1} \delimsize\vert{#2}
}
\DeclarePairedDelimiterX{\dyad}[2]{\lvert}{\vert}{
	{#1} \delimsize\rangle \delimsize \langle{#2}
}

\begin{document}
	\title{Limitations of an approximative phase-space description in strong-field quantum optics}
	
	\author{Rasmus Vesterager Gothelf}
	\author{Lars Bojer Madsen}
	\author{Christian Saugbjerg Lange} 
	
	\affiliation{Department of Physics and Astronomy, Aarhus University, Ny Munkegade 120, DK-8000 Aarhus C, Denmark}
	\date{\today}
	\begin{abstract}
		In recent years, strong-field processes such as high-order harmonic generation (HHG) and above-threshold ionization driven by nonclassical states of light have become an increasingly popular field of study. The theoretical modeling of these processes often applies an approximate phase-space expansion of the nonclassical driving field in terms of coherent states, which has been shown to accurately predict the harmonic spectrum. However, its accuracy for the computation of quantum optical observables like the degree of squeezing and photon statistics has not been thoroughly considered. 
		In this work, we introduce this approximative phase-space description and discuss its accuracy, and we find that it mischaracterizes the quantum optical properties of the driving laser by making it an incoherent mixture of classical states. We further show that this error in the driving field description maps onto the light emitted from HHG, as neither sub-Poissonian photon statistics nor quadrature squeezing below vacuum fluctuations can be captured by the approximative phase-space description. 
		Lastly, to benchmark the approximative phase-space description, we consider the quantum HHG from a one-band model, which yields an exact analytical solution. 
		Using the approximative phase-space representation with this specific model, we find a small quantitative error in the quadrature variance of the emitted field that scales with pulse duration and emitter density. 
		Our results show that using this approximative phase-space description can mischaracterize quantum optical observables. Attributing physical meaning to such results should therefore be accompanied by a quantitative analysis of the error.  
	\end{abstract}
	
	\maketitle
	
	\section{Introduction}
	
	Strong-field quantum optics is a field in rapid development. Following recent work on the quantum optical aspects of strong-field processes such as high-harmonic generation (HHG) \cite{Gorlach2020,Lewenstein2021,Lange2024a,Lange2025a,Lange2025b,Yi2024,Stammer2022,Theidel2024,Lange2025c} alongside advancements in the production of high-intensity quantum fields, called bright-squeezed vacuum (BSV) \cite{Manceau2019,Agafonov2010,Iskhakov2012,Perez2014,Chekhova2015,Finger2015,Sharapova2020}, studies of driving HHG with quantum states of light have received much interest. Specifically, the effects of squeezed driving fields are of interest both theoretically \cite{Gorlach2023,Gothelf2025,Stammer2024,EvenTzur2023,Wang2025,Petrovic2026} and in various strong-field experiments \cite{Rasputnyi2024,Heimerl2024,Heimerl2025,Sennary2026}.

	One of the central theoretical methods applied in these studies using nonclassical driving fields is a coherent-state expansion \cite{Gorlach2023}, where the quantum state of a general driving field is expressed in terms of coherent states, motivated by previous works on coherently driven strong-field processes \cite{Gorlach2020}.
	Such an expansion maps the case of a general driving field onto the well-known case of a coherent driving field.
	This method has proven to effectively predict the HHG spectra for nonclassical driving fields \cite{Gorlach2023,Gothelf2025,Stammer2024,EvenTzur2023,Wang2025}.

	However, for this method to be computationally feasible in most cases, an approximation is applied \cite{Gorlach2023,Stammer2025b,Rivera-Dean2025a,Gothelf2025,Wang2025}, which we call the approximative positive $P$ (APP) representation \cite{Gothelf2025}. This approximation assumes the coherent state expansion to be diagonal in a multidimensional phase space and the associated distribution function to be positive and smooth.

	The APP representation has proved useful in computing the spectra from HHG \cite{Gorlach2023,Gothelf2025,Wang2025}, ATI \cite{Wang2023,Liu2025,FangLiu2023,Rivera-Dean2025c,Lyu2025}, and for interpreting results of experiments \cite{Heimerl2024} when driving with quantum states of light. However, the accuracy of the APP representation has not been verified for all relevant observables. Although the APP representation accurately predicts HHG spectra \cite{Wang2025,Gothelf2025,Gonzalez-Monge2025}, its accuracy for quantum optical observables such as photon statistics and squeezing \cite{Gothelf2025,Wang2025,Stammer2024}, as well as the electron-photon joint energy spectrum \cite{Mao2025}, is questioned. 
	
	As more interest in the field develops and this approximation finds more use, it is therefore relevant to analyze the accuracy of the APP representation, in particular its predictions of quantum optical observables from HHG, such that results obtained through it are interpreted with fitting skepticism.

	In this work, we demonstrate how the APP representation cannot capture the quantum optical properties of quadrature squeezing below vacuum fluctuations and sub-Poissonian photon statistics for both the driving field and the emitted HHG field, whether these properties are physically present or not, when considering an electronic model in which the dipole correlations are neglected. We quantify this error introduced by the APP representation for a coherently driven one-band model of a solid, which yields an analytic solution.

	The paper is organized as follows: In Sec. \ref{Sec: Theory}, to aid the discussion of the APP representation and for completeness, we present the general derivation for HHG driven by quantum light using phase-space expansions, and we discuss the choice of representation, including an introduction and discussion of the APP representation. In Sec. \ref{Sec: Limitations}, we show that in the absence of dipole correlations in the electronic system, no quadrature squeezing below vacuum fluctuations or sub-Poissonian photon statistics can be found. We also analyze the error introduced by the APP in the driving field and in the HHG light emitted from a one-dimensional single-band solid driven by coherent light. Lastly, in Sec. \ref{Sec: conclusion} we give a conclusion and an outlook. In Appendices \ref{App: wigner} and \ref{App: details_oneband}, we show that the Wigner function is nonnegative when calculated using the APP, and we provide details on calculations for the one-band model.

	Atomic units ($\hbar=m_e=4\pi\epsilon_0=e=1$) are used throughout the paper unless otherwise stated.

	\section{Theory} \label{Sec: Theory}
	We first present a derivation of the generated light from HHG with a general driving field, as also found in Refs. \cite{Gorlach2023,Gothelf2025}. Here, we discuss the different choices of representation and formally introduce the APP representation.

	
	\subsection{General formulation} \label{Sec: Formulation}
	We consider the Hamiltonian for a general electronic system driven by a quantum field
	\begin{align}
		\hat{H} = \frac{1}{2}\sum_j\big(\vo{p}_j+\vo{A}\big)^2 + \hat{U} + \hat{H}_\text{F}, \label{Eq: general_hamiltonian}
	\end{align}
	where $\vo{p}_j$ is the momentum of electron $j$, $\vo{A} = \sum_{\vec{k}\sigma} \frac{g_0}{\sqrt{\omega_{\vec{k}}}} \big(\hat{a}_{\vec{k}\sigma}\vec{e}_{\sigma}+\hat{a}_{\vec{k}\sigma}^\dag\vec{e}_{\sigma}^*\big)$ is the quantized vector potential in the dipole approximation with $\omega_{\vec{k}}$ being the frequency corresponding to the wavevector $\vec{k}$, and $\vec{e}_\sigma$ being the polarization unit vector corresponding to polarization $\sigma$. The annihilation (creation) operator of the spectral mode $(\vec{k},\sigma)$ is denoted $\hat{a}_{\vec{k},\sigma}$ ($\hat{a}_{\vec{k},\sigma}^\dag$), and $g_0=\sqrt{2\pi/V}$ is the coupling constant determined by the quantization volume $V$. Furthermore, $\hat{U}$ is the electron potential and $\hat{H}_\text{F}=\sum_{\vec{k}\sigma}\omega_{\vec{k}}\hat{a}_{\vec{k},\sigma}^\dag\hat{a}_{\vec{k},\sigma}$ is the free-field Hamiltonian.

	First, we solve the Schrödinger equation for the Hamiltonian of Eq. (\ref{Eq: general_hamiltonian}) with a coherent driving field as in Refs. \cite{Gorlach2020,Lange2024a}. This is done by assuming the initial condition to be in the electronic groundstate, $\ket{\phi_i}$, with a coherent state $\ket{\alpha}$ in the laser mode $(\vec{k}_L,\sigma_L)$ and with the vacuum state $\ket{0}$ in all other photonic modes. We thus write the initial state as $\ket{\psi^\alpha(0)}=\ket{\phi_i}\otimes\ket{\alpha}\bigotimes_{(\vec{k},\sigma)\neq(\vec{k}_L,\sigma_L)}\ket{0}$, letting the initial state be prepared at time $t=0$ for simplicity. In Ref. \cite{Gonzalez-Monge2025}, this initial state was considered in terms of temporal mode quantum states.

	Switching to a displaced interaction picture with $\ket{\tilde{\psi}^\alpha(t)}=\hat{\mathcal{U}}_{\text{sc}}^{\alpha\dag}(t) \hat{D}^\dag(\alpha) \hat{\mathcal{U}}_{F}^\dag(t)\ket{\psi(t)}$, where $\hat{\mathcal{U}}_{F}(t)$ is the time-evolution operator of $\hat{H}_F$, $\hat{D}(\alpha)$ is the displacement operator in the laser mode, and $\hat{\mathcal{U}}^{\alpha}_{\text{sc}}(t)$ is the semiclassical time-evolution operator for the system driven by the classical laser $\vec{A}^\alpha_\text{cl}(t)=\frac{g_0}{\sqrt{\omega_{\vec{k}_L}}} \big(\alpha e^{-i\omega_L t} \vec{e}_{\sigma_L}+\alpha^* e^{i\omega_L t} \vec{e}_{\sigma_L}^*\big)$, we obtain the time-dependent Schrödinger equation (neglecting an $\vo{A}^2$ term \cite{Gorlach2020,Lange2024a})
	\begin{align}
		i\frac{\partial}{\partial t}\ket{\tilde{\psi}^\alpha(t)} = \vo{A}(t) \cdot \vo{j}^\alpha_\text{sc,I}(t) \ket{\tilde{\psi}^\alpha(t)}, \label{Eq: EOM_general}
	\end{align}
	where $\vo{j}^\alpha_\text{sc,I}(t) = \hat{\mathcal{U}}^{\alpha\:\dag}_{\text{sc}}(t) \sum_j\big[\vo{p}_j+\vec{A}^\alpha_\text{cl}(t)\big]\hat{\mathcal{U}}^\alpha_{\text{sc}}(t)$ is the semiclassical current in the interaction picture, and where the quantized vector potential has gained a time dependence via $\vo{A}(t)=\hat{\mathcal{U}}_{F}^\dag(t)\vo{A}\hat{\mathcal{U}}_{F}(t)$.

	To solve Eq. (\ref{Eq: EOM_general}) one typically expands the combined light-matter state in the set of electronic eigenstates $\{\ket{\phi_n}\}$ so that $\ket{\tilde{\psi}^\alpha(t)} = \sum_n \ket{\phi_n}\otimes\ket{\chi_n^\alpha(t)}$, leaving the photonic wave packets $\{\ket{\chi_n^\alpha(t)}\}$ unknown. Projecting onto the electronic state $\bra{\phi_n}$, an equation of motion for the photonic wave packets is obtained
	\begin{align}
		i \frac{\partial}{\partial t} \ket{\chi_m^\alpha(t)} = \vo{A}(t) \cdot \sum_n \vec{j}^\alpha_{m,n}(t) \ket{\chi_n^\alpha(t)}, 
		\label{Eq: EOM_photonic}
	\end{align}
	where $\vec{j}^\alpha_{m,n}(t) = \mel{\phi_m}{\vo{j}_{\text{sc,I}}(t)}{\phi_n}$ are the so-called transition currents, as they connect, in general, two different electronic states. The off-diagonal transition currents account for the time correlations in the dipole operator as seen in Refs. \cite{Lange2025c,Lange2025a,Stammer2024a,Stammer2025}. 
	
	Solving Eq. (\ref{Eq: EOM_photonic}) specifies the state of the entire system as $\ket{\psi^\alpha(t)}=\hat{\mathcal{U}}_F(t)\hat{D}(\alpha)\hat{\mathcal{U}}_{\text{sc}}^\alpha(t)\sum_n\ket{\phi_n}\otimes\ket{\chi_n^\alpha(t)}$. An analytic solution is given in Sec. \ref{Sec: analytic_solution} for a system without dipole correlations, i.e., where $\vec{j}^\alpha_{m,n}(t)=0$ for $m\neq n$.

	We now turn our attention to a general driving field.
	The initial state of the density operator of the system is written as 
	\begin{align}
		\hat{\rho}(0)= \dyad{\phi_i}{\phi_i}\otimes \int d\mu \frac{P(\alpha,\beta)}{\inner{\beta^*}{\alpha}}\dyad{\alpha}{\beta} \bigotimes_{(\vec{k},\sigma)\neq(\vec{k}_L,\sigma_L)} \dyad{0}{0}, \label{Eq. rho_initial}
	\end{align}
	where we have expanded the driving field mode in terms of coherent states \cite{Gorlach2023} using a generalized $P$ representation \cite{Walls2008,Drummond1980,Gothelf2025}. Due to the overcompleteness of the coherent states, this expansion is not unique \cite{Walls2008}. Hence, for generality of notation, the integration measure $d\mu$ is not specified, thereby retaining this degree of freedom. Once such a measure is chosen, the state of the driving laser is described by a unique $P$ function (if one exists).

	The density operator of the system has to fulfill the von Neumann equation of motion $i\partial_t\hat{\rho}(t)=[\hat{H},\hat{\rho}(t)]$ with the Hamiltonian given in Eq. (\ref{Eq: general_hamiltonian}) along with the initial condition specified in Eq. (\ref{Eq. rho_initial}). It is then simple to check that the density operator
	\begin{align}
		\hat{\rho}(t) = \int d\mu \frac{P(\alpha,\beta)}{\inner{\beta^*}{\alpha}} \dyad{\psi^\alpha(t)}{\psi^{\beta^*}(t)} \label{Eq: density_operator_general_solution}
	\end{align}
	indeed fulfills both, as $i\partial_t\ket{\psi^\alpha(t)}=\hat{H}\ket{\psi^\alpha(t)}$. Equation (\ref{Eq: density_operator_general_solution}) is thus the general state of a system driven by a single-mode quantum field.

	The strength of this coherent phase-space expansion [Eq. (\ref{Eq: density_operator_general_solution})] is apparent. By framing the problem in terms of coherent states, we map it onto a semiclassical problem: First, the semiclassical transition currents $\vec{j}^\alpha_{m,n}(t)$ are computed using semiclassical methods. The emitted photonic state for a system driven by a coherent state is then found using Eq. (\ref{Eq: EOM_photonic}). Lastly, the quantum state of the driving field is accounted for by integrating the coherent solutions with the distribution function $P(\alpha,\beta)$, as given by Eq. (\ref{Eq: density_operator_general_solution}).

	In the context of strong-field quantum optics, the utilized representations are typically the Glauber-Sudarshan (GS) and the positive $P$ (PP) representation.

	The GS representation chooses $d\mu=d^2\alpha d^2\beta \delta^{(2)}(\alpha-\beta^*)$ and the state of the system becomes
	\begin{align}
		\hat{\rho}(t) = \int d^2\alpha P^{(\text{GS})}(\alpha) \dyad{\psi^\alpha(t)}{\psi^{\alpha}(t)}, \label{Eq: density_operator_GS}
	\end{align}
	with a driving-field-specific distribution $P^{(\text{GS})}(\alpha)$. This representation is diagonal in the coherent states; however, the distribution function can be highly singular for nonclassical fields. Indeed, a way to define a nonclassical field is a state for which the GS distribution is negative or more singular than a delta function \cite{Gerry2004}. For this reason, the GS representation is not practical for computation when driving with nonclassical fields such as BSV or Fock states.

	On the other hand, the PP representation chooses $d\mu = d^2\alpha d^2\beta$, yielding the expression for the system state
	\begin{align}
		\hat{\rho}(t) = \int  d^2\alpha d^2\beta \frac{P^{(\text{PP})}(\alpha,\beta)}{\inner{\beta^*}{\alpha}} \dyad{\psi^\alpha(t)}{\psi^{\beta^*}(t)}, 
		\label{Eq: density_operator_PP}
	\end{align}
	with a driving-field-specific distribution $P^{(\text{PP})}(\alpha,\beta)$. As opposed to the GS representation, the PP distribution is positive and smooth for all states. However, it is not diagonal, in the sense that the outer products $\dyad{\psi^\alpha(t)}{\psi^{\beta^*}(t)}$ in the integrand consist of two separate states, which again leaves this approximation impractical for computation. 
	This impracticality stems from the high dimensionality of the integral, and from the nontriviality of the inner product of $\ket{\psi^\alpha(t)}$ and $\ket{\psi^{\beta^*}(t)}$: When calculating the matrix element $\mel{\psi^{\beta^*}(t)}{\hat{O}_{\vec{k},\sigma}}{\psi^\alpha(t)}$ of a single-mode operator $\hat{O}_{\vec{k}\sigma}$, the inner products of the two states in the remaining modes are not trivially 1 and all photonic modes would therefore need to be considered in the calculation. Though an expression for this overlap has been considered in the continuum limit of $g_0\rightarrow0$ in Ref. \cite{Gonzalez-Monge2025}, it is, in general, difficult to address and numerically demanding.

	As is apparent, though this coherent phase-space expansion is attractive for its relation to coherently driven (and therefore also semiclassical) dynamics, both representations prohibit explicit computation of observables for certain driving fields of interest, such as BSV. We stress, however, that if computation is possible in either representation, the presented theory is exact.

	\subsection{The approximative positive $P$ (APP) representation} \label{Sec: APP}
	
	To overcome the challenges presented above, an approximate representation was first introduced in Ref. \cite{Gorlach2023}. This approximation uses the relation between the positive $P$ distribution function and the Husimi function $Q(\alpha)$ \cite{Gerry2004,Drummond1980} 
	\begin{align}
		P^{(\text{PP})}(\alpha,\beta)=\frac{1}{4\pi}e^{-\frac{|\alpha-\beta^*|^2}{4}} Q\bigg(\frac{\alpha+\beta^*}{2}\bigg), 
		\label{Eq: P_PP_Q_relation}
	\end{align}
	where the Husimi function is defined as
	\begin{align}
		Q(\alpha) = \frac{1}{\pi}\mel{\alpha}{\hat{\rho}_L}{\alpha},
		\label{Eq: Q_def}
	\end{align}
	for the density operator of the driving field $\hat{\rho}_L$.
	
	To make the PP function diagonal, the Gaussian function in Eq. (\ref{Eq: P_PP_Q_relation}) is approximated as a delta function.
	\begin{align}
		P^{(\text{PP})}(\alpha,\beta)\approx 
		P^{(\text{APP})}(\alpha,\beta) =
		\delta^{(2)}(\alpha-\beta^*) Q\bigg(\frac{\alpha+\beta^*}{2}\bigg), 
		\label{Eq: APP_P_to_Q}
	\end{align}
	where $\delta^{(2)}$ denotes the two-dimensional Dirac delta distribution. We call the function in Eq. (\ref{Eq: APP_P_to_Q}) the approximative positive $P$ (APP) function. 
	
	By substituting Eq. (\ref{Eq: APP_P_to_Q}) into Eq. (\ref{Eq: density_operator_PP}), we get an approximate representation, where the system state is expressed as
	\begin{align}
		\hat{\rho}(t) \approx \hat{\rho}^{(\text{APP})}(t) = \int d^2\alpha Q(\alpha) \dyad{\psi^\alpha(t)}{\psi^{\alpha}(t)}, \label{Eq: density_operator_APP}
	\end{align}
	which we call the APP representation of the system state. We note that the state of the driving laser and the laser parameters enter Eq. (\ref{Eq: density_operator_APP}) through the field-specific Husimi function, $Q(\alpha)$.

	The Husimi function exists and is smooth and positive for all driving fields \cite{Gerry2004,Walls2008}. Thus, the APP representation is both diagonal and has a well-defined distribution function, overcoming the limitations of the GS and PP representations, and making it convenient for numerical computation of observables.

	The approximation of Eq. (\ref{Eq: APP_P_to_Q}) has been reasoned in different ways. In the original introduction in Ref. \cite{Gorlach2023}, it is argued under a coordinate transform to field amplitudes $\epsilon_\alpha= 2g_0\alpha$, which are held constant while taking the limit $g_0\rightarrow 0$, corresponding to assuming an infinite quantization volume. 
	In Ref. \cite{Stammer2025b}, this limit is further justified by keeping the product of $g_0$ and the number of emitters constant while taking the limit of $g_0\rightarrow 0$. 
	In Ref. \cite{Gonzalez-Monge2025}, it is introduced as a lowest order expansion of Eq. (\ref{Eq: density_operator_PP}) in a relative coordinate $\frac{\alpha-\beta}{\alpha+\beta}$, and
	in Ref. \cite{Gothelf2025}, it was justified under geometric arguments for the specific positive $P$ distributions in question.
	In the present work, we do not further consider the justification for using the APP, but instead focus on its accuracy for computing quantum optical observables.

	At this point, we may discuss some physical implications of the APP representation. Firstly, we note that though the applicability of the APP representation might in some works be argued under properties of the HHG target, such as having a large amount of HHG emitters in the target \cite{Stammer2025b,Rivera-Dean2025a}, the approximation is essentially an approximation on the driving field state, as the initial condition of the system becomes
	\begin{align}
		\hat{\rho}^{(\text{APP})}(0) = \dyad{\phi_i}{\phi_i}\otimes\int d^2\alpha Q(\alpha)\dyad{\alpha}{\alpha}\bigotimes_{(\vec{k},\sigma)\neq(\vec{k}_L,\sigma_L)}\dyad{0}{0},
	\end{align}
	meaning that the state of the driving field is $\hat{\rho}_L^{(\text{APP})}=\int d^2\alpha Q(\alpha)\dyad{\alpha}{\alpha}$. We note that since the laser field state has been put on a diagonal form with a positive and smooth distribution function, we have effectively assumed the laser to be an incoherent mixture of coherent states, i.e., a classical field, independent of the nature of the driving field, classical or nonclassical. In Sec. \ref{Sec: driving} we show examples of how this mixed-state description leads to a mischaracterization of the quantum optical aspects of the driving field. 
	
	\begin{figure}
		\includegraphics[width=\linewidth]{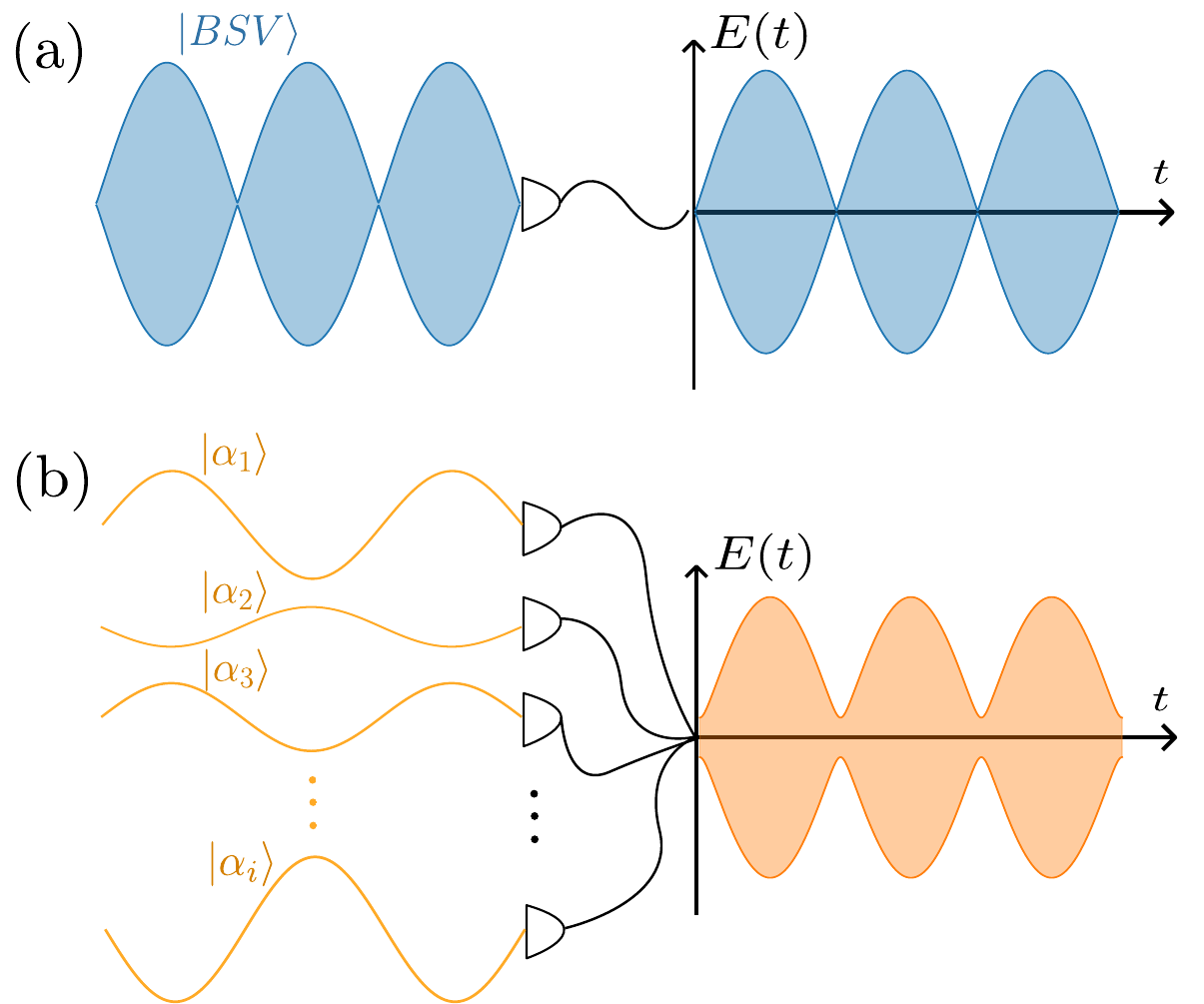}
		\caption{Illustration of (a) a pulse of BSV and (b) multiple shots of a classical laser with stochastic amplitude and phase. Comparing the electric fields that would be measured for both fields, both have no mean field and an uncertainty that oscillates in time. The uncertainties are similar when intensity is large, but the minimum uncertainty in the stochastic laser (b) is bounded from below by $1/4$, whereas the BSV field (a) is not. As discussed in the main text, using the APP representation is equivalent to approximating (a) as (b).}
		\label{Fig: Stochastic_vs_BSV}
	\end{figure}
	
	Nevertheless, while the APP representation may fail to capture quantum optical aspects of the driving field, one might expect it to still capture the relevant features of the generated field \cite{Stammer2025b}. It has already been seen that the APP representation captures the HHG spectrum well \cite{Wang2025,Gothelf2025}, validating results studying HHG spectra \cite{Gorlach2023,Gothelf2025}. 
	
	In Sec. \ref{Sec: General}, we show that quantum optical observables, such as quadrature squeezing and sub-Poissonian photon statistics, of the generated field are not captured if present.
	Indeed, one might expect this already from Eq. (\ref{Eq: density_operator_APP}). If an operator $\hat{O}$, for which the expectation value for a coherent driving field is denoted $\mean{\hat{O}}_\alpha=\mel{\psi^\alpha}{\hat{O}}{\psi^\alpha}$, is considered in the APP representation, its expectation value for a general field would be
	\begin{align}
		\mean{\hat{O}} = \int d^2\alpha Q(\alpha) \mean{\hat{O}}_\alpha, 
		\label{Eq: APP_expectation_value}
	\end{align}
	where $Q(\alpha)$ represents the distribution of the driving field state in terms of coherent states, which is a classical mixture due to the nonnegativity and smoothness of the Husimi function, and is determined from Eq. (\ref{Eq: Q_def}).

	The measurement described in Eq. (\ref{Eq: APP_expectation_value}) can be reproduced in an experiment with a classical (coherent) laser:
	We could let the phase and amplitude of a coherent driver (captured by the complex number $\alpha$) be chosen stochastically in each shot of the experiment according to a classical probability distribution $P_{\text{cl}}(\alpha)$. The multi-shot average of the studied observable $\hat{O}$ would, by classical probability theory, be
	\begin{align}
		\mean{\hat{O}} = \int d^2\alpha P_{\text{cl}}(\alpha) \mean{\hat{O}}_\alpha. 
		\label{Eq: stochastic_expectation_value}
	\end{align}
	As the Husimi function $Q(\alpha)$ is a smooth and positive function, i.e., a classical distribution, we could choose $P_{\text{cl}}(\alpha)=Q(\alpha)$, and Eqs. (\ref{Eq: APP_expectation_value}) and (\ref{Eq: stochastic_expectation_value}) would be equal. Thus, we see that in a multi-shot experiment, this approximation for a quantum driving field is indistinguishable from a classical driving field with stochastically determined parameters. As a classically driven experiment would not exhibit quantum optical features, we do not expect the APP representation to capture these features either due to this equivalence.
	
	In Fig. \ref{Fig: Stochastic_vs_BSV}, an illustration of a BSV field [Fig. \ref{Fig: Stochastic_vs_BSV}(a)] and a stochastic classical laser [Fig. \ref{Fig: Stochastic_vs_BSV}(b)] is depicted. The uncertainties in the electric field in multi-shot experiments are similar for the two fields, though the classical field cannot exhibit a minimum variance below $1/4$. In using the APP, the BSV field in Fig. \ref{Fig: Stochastic_vs_BSV}(a) is approximated as the classical stochastic field in Fig. \ref{Fig: Stochastic_vs_BSV}(b).

	\subsection{Analytical solution for coherently driven systems with no dipole correlations} \label{Sec: analytic_solution}
	For use in Secs. \ref{Sec: General} and \ref{Sec: one-band}, we present the analytic solution of Eq. (\ref{Eq: EOM_photonic}) for a system where the current operator is diagonal, i.e., if $\vec{j}^\alpha_{m,n}(t)\propto \delta_{m,n}$. Physically, this condition states that there are no dipole correlations \cite{Lange2024a,Lange2025c}. If we assume that this is the case, the solution is a multimode coherent state \cite{Gerry2004,Scully1997,Gorlach2020,Lange2024a}
	\begin{align}
		\ket{\chi_m^\alpha(t)} = \delta_{m,i} \bigotimes_{\vec{k},\sigma}D\big[\gamma^\alpha_{\vec{k},\sigma}(t)\big] \ket{0}, \label{Eq: coherent_solution}
	\end{align} 
	where the subscript $i$ denotes the initial electronic state, and where
	\begin{align}
		\gamma^\alpha_{\vec{k},\sigma}(t) = -i\frac{g_0}{\sqrt{\omega_{\vec{k}}}}\int_0^t \vec{j}^\alpha_{i,i}(t')\cdot\vec{e}_{\sigma}e^{i\omega_{\vec{k}}t'}dt',
		\label{Eq: gamma_expression}
	\end{align}
	is the coherent state parameter for the emitted light in mode $(\vec{k},\sigma)$. Again, the driving field is assumed to turn on at time $t=0$ for simplicity, and therefore the lower bound on the integral in Eq. (\ref{Eq: gamma_expression}) is $0$.

	For some systems, such as, e.g., a one-band model of a solid \cite{Lange2024a,Gothelf2025}, which is considered in Sec. \ref{Sec: one-band}, Eqs. (\ref{Eq: coherent_solution}) and (\ref{Eq: gamma_expression}) are exact, but we emphasize that for most systems, such as atomic systems, multi-band models, or models with electron correlations, this is not the case \cite{Gorlach2020,Lange2024a,Lange2025b}, though it is often assumed for simplicity \cite{Gorlach2023,Stammer2024a,Stammer2025b,Stammer2024,Rivera-Dean2025a,Gonzalez-Monge2025,Wang2025}.

	\section{Accuracy of the APP representation} \label{Sec: Limitations}
	
	We now present an analysis and discussion of the accuracy of the APP representation, by first showing a general proof that quantum optical features are not present under the APP representation when driving a system without dipole correlations with nonclassical light in Sec. \ref{Sec: General}. For a quantitative benchmark, we consider the error introduced by the APP representation in (i) the driving field in Sec. \ref{Sec: driving} and (ii) in an example of the field generated from HHG in a single-band solid driven by coherent light in Sec. \ref{Sec: one-band}.

	For convenience, we state mathematically the observables of interest: The spectrum is given by \cite{Gorlach2020,Gorlach2023,Lange2024a,Gothelf2025}
	\begin{align}
		S(\omega_{\vec{k}})\propto \sum_\sigma \mean{\hat{a}_{\vec{k},\sigma}^\dag \hat{a}_{\vec{k},\sigma}}. \label{Eq: spectrum_general}
	\end{align}
	As stated previously, multiple previous works have found that the spectrum is well captured by the APP representation \cite{Gothelf2025,Wang2025,Gonzalez-Monge2025}, which we again find in Sec. \ref{Sec: one-band}.
	
	Quadrature squeezing is quantified by the variance of the quadrature with angle $\theta$, defined as $\hat{X}_{\vec{k},\sigma}(\theta)=\frac{1}{2}(\hat{a}_{\vec{k},\sigma}e^{-i\theta}+\hat{a}_{\vec{k},\sigma}^\dag e^{i\theta})$, and is given as
	\begin{align}
		&\mean{\Delta \hat{X}_{\vec{k},\sigma}^2(\theta)} = 
		\frac{1}{4}\bigg[
		1+
		2\Big(\mean{\hat{a}_{\vec{k},\sigma}^{\dag}\hat{a}_{\vec{k},\sigma}}-\mean{\hat{a}^\dag_{\vec{k},\sigma}}\mean{\hat{a}_{\vec{k},\sigma}}\Big) \notag \\
		&+ e^{-2i\theta}\Big(\mean{\hat{a}_{\vec{k},\sigma}^2}-\mean{\hat{a}_{\vec{k},\sigma}}^2\Big) 
		+ e^{2i\theta}\Big(\mean{\hat{a}_{\vec{k},\sigma}^{\dag\:2}}-\mean{\hat{a}^\dag_{\vec{k},\sigma}}^2\Big) 
		\bigg], \label{Eq: squeezing_operator}
	\end{align}
	with a minimum variance below 1/4 signifying a quantum optical squeezed state, as it has a lower variance in one quadrature than the vacuum state.

	Lastly, the photon statistics are quantified by the second-order normalized coherence function
	\begin{align}
		g_{\vec{k},\sigma}^{(2)}(0) = \frac{\mean{\hat{a}_{\vec{k},\sigma}^\dag \hat{a}_{\vec{k},\sigma}^\dag \hat{a}_{\vec{k},\sigma}\hat{a}_{\vec{k},\sigma}}}{\mean{\hat{a}_{\vec{k},\sigma}^\dag\hat{a}_{\vec{k},\sigma}}^2},
		\label{Eq: g2_form}
	\end{align}
	where $g_{\vec{k},\sigma}^{(2)}(0)=1$ signifies Poissonian statistics, found in, e.g., coherent states, while $g_{\vec{k},\sigma}^{(2)}(0)<1$ signifies sub-Poissonian statistics, found in, e.g., Fock states and $g_{\vec{k},\sigma}^{(2)}(0)>1$ signifies super-Poissonian statistics, found in, e.g., thermal and BSV states \cite{Gerry2004}.
	
	For clarity, we distinguish between exact expectation values and expectation values calculated using the APP representation by denoting the latter by $\mean{\cdot}^{(\text{APP})}$.

	\subsection{General statements} \label{Sec: General}
	We start by proving that squeezing below vacuum fluctuations and sub-Poissonian photon statistics cannot be found for nonclassically driven HHG when using the APP representation 
	under the assumption that driving the system with a coherent state generates a coherent state from HHG. As discussed in Sec. \ref{Sec: analytic_solution}, although the coherent-in-coherent-out assumption is not exact for most electronic systems, it is widely adopted \cite{Gorlach2023,Stammer2024a,Stammer2025b,Stammer2024,Rivera-Dean2025a,Gonzalez-Monge2025,Wang2025} and therefore relevant to the present work.

	Together, the APP representation and the coherent-in-coherent-out assumption yield expressions via Eqs. (\ref{Eq: APP_expectation_value}) and (\ref{Eq: gamma_expression}) for the relevant expectation values [using the shorthand notation $n=(\vec{k},\sigma)$ for a general mode] that read
	\begin{subequations}
		\begin{align}
			\mean{\hat{a}_n^\dag \hat{a}_n}^{(\text{APP})} &= \int d^2 \alpha Q(\alpha) \abs{\gamma_n^\alpha}^2 \\
			\big[\mean{\hat{a}_n^2}^{(\text{APP})}\big]^* = \mean{\hat{a}_n^2}^{(\text{APP})} &= \int d^2 \alpha Q(\alpha) \gamma_n^{\alpha\:2} \\
			\big[\mean{\hat{a}}^{(\text{APP})}\big]^* = \mean{\hat{a}_n}^{(\text{APP})} &= \int d^2 \alpha Q(\alpha) \gamma_n^\alpha \\
			\mean{\hat{a}_n^{\dag\:2} \hat{a}_n^2}^{(\text{APP})} &= \int d^2 \alpha Q(\alpha) \abs{\gamma_n^\alpha}^4.
		\end{align}
		\label{Eq: general_APP_a_values}
	\end{subequations}

	First, we consider quadrature squeezing. From Eq. (\ref{Eq: squeezing_operator}), it is apparent that the minimum quadrature variance is given by 
	\begin{align}
		\min_{\theta\in[0,\pi)}\mean{\Delta \hat{X}_n^2(\theta)} = \frac{1}{4} + \frac{1}{2}
		\bigg[
		\Big(\mean{\hat{a}_n^\dag \hat{a}_n} - \mean{\hat{a}_n}\mean{\hat{a}_n^\dag} \Big) 
		\notag \\
		- \Big|\mean{\hat{a}_n^2} - \mean{\hat{a}_n}^2 \Big|
		\bigg].
	\end{align}
	Hence, the condition for squeezing below vacuum fluctuations, which signifies a quantum optical squeezed state, is $\Big|\mean{\hat{a}_n^2} - \mean{\hat{a}_n}^2 \Big| > \mean{\hat{a}_n^\dag \hat{a}_n} - \mean{\hat{a}_n}\mean{\hat{a}_n^\dag}$. However in the APP representation using Eq. (\ref{Eq: general_APP_a_values}) we find that
	\begin{align}
		&\Big|\mean{\hat{a}_n^2}^{(\text{APP})} - \mean{\hat{a}_n}^{(\text{APP})\:2} \Big| 
		\notag \\
		&= \Bigg|
		\int d^2\alpha Q(\alpha)  \bigg[\gamma_n^{\alpha\:2} - \int d^2\beta Q(\beta)\gamma_n^\beta\bigg]^2
		\Bigg|
		\notag \\
		&\leq 
		\int d^2\alpha   \bigg| Q(\alpha) \bigg[\gamma_n^{\alpha\:2} - \int d^2\beta Q(\beta)\gamma_n^\beta\bigg]^2 \bigg|,
		\label{Eq: general_statement_squeezing_first_inequality}
	\end{align}
	where the inequality is a standard result of integrals. We can use the fact that $Q(\alpha)$ is a nonnegative function to pull it outside the absolute value in Eq. (\ref{Eq: general_statement_squeezing_first_inequality}) and obtain
	\begin{align}
		&\Big|\mean{\hat{a}_n^2}^{(\text{APP})} - \mean{\hat{a}_n}^{(\text{APP})\:2} \Big| 
		\notag \\
		&\leq 
		\int d^2\alpha   Q(\alpha) \bigg|\gamma_n^{\alpha\:2} - \int d^2\beta Q(\beta)\gamma_n^\beta\bigg|^2 
		\notag \\
		&=
		\int d^2\alpha   Q(\alpha) \big|\gamma_n^{\alpha}\big|^2 +  \bigg|\int d^2\beta Q(\beta)\gamma_n^\beta\bigg|^2 
		\notag \\
		&\qquad- 2\text{Re}\bigg[\int d^2\alpha   Q(\alpha) \gamma_n^{\alpha} \int d^2\beta   Q(\beta) \gamma_n^{\beta\:*}\bigg]
		\notag \\
		&=
		\int d^2\alpha   Q(\alpha) \big|\gamma_n^{\alpha}\big|^2 -  \bigg|\int d^2\beta Q(\beta)\gamma_n^\beta\bigg|^2 
		\notag \\
		&=
		\mean{\hat{a}_n^\dag \hat{a}_n}^{(\text{APP})} - \mean{\hat{a}_n}^{(\text{APP})}\mean{\hat{a}_n^\dag}^{(\text{APP})},
	\end{align}
	which shows that $\Big|\mean{\hat{a}_n^2}^{(\text{APP})} - \mean{\hat{a}_n}^{(\text{APP})\:2} \Big| \leq \mean{\hat{a}_n^\dag \hat{a}_n}^{(\text{APP})} - \mean{\hat{a}_n}^{(\text{APP})}\mean{\hat{a}_n^\dag}^{(\text{APP})}$. Thus, we have proved that from the expectation values in Eq. (\ref{Eq: general_APP_a_values}), squeezing below vacuum fluctuations is unattainable in the APP representation, independent of the quantum optical nature of the driving field.

	In contrast, works have already predicted small but definite squeezing in the emitted field without using the APP representation for both a coherent driving field \cite{Gorlach2020,Lange2024a,Lange2025b,Stammer2022} and for a displaced squeezed driving field \cite{EvenTzur2024}, which would not be captured using the APP.

	Using the APP representation, squeezing below vacuum fluctuations cannot be found, and one can therefore not discern whether the absence hereof is physical or a consequence of using the APP representation. Thus, physical meaning cannot be attributed to the absence of squeezing below vacuum fluctuations when using the APP representation.

	We can likewise consider the photon statistics. Rewriting Eq. (\ref{Eq: g2_form}) and evaluating in the APP representation, we see that
	\begin{align}
		g^{(2)(\text{APP})}(0) = 1+ \frac{\mean{\hat{a}_n^{\dag\:2} \hat{a}_n^2}^{(\text{APP})}-\mean{\hat{a}_n^\dag \hat{a}_n}^{(\text{APP})\:2}}{\mean{\hat{a}_n^\dag \hat{a}_n}^{(\text{APP})\:2}} 
		\notag \\
		= 1 + \frac{\int d^2\alpha Q(\alpha)\big[
			\abs{\gamma_n^{\alpha}}^2 - 	\int d^2\beta Q(\beta)\abs{\gamma_n^{\beta}}^2
			\big]^2}{\mean{\hat{a}_n^\dag \hat{a}_n}^{(\text{APP})\:2}},
	\end{align}
	where the numerator is the integral over a product of two nonnegative functions [the Husimi function $Q(\alpha)$ and the square of a real-valued function], which implies that $g^{(2)(\text{APP})}(0) \geq 1$, meaning that sub-Poissonian photon statistics cannot be predicted by the APP representation.
	Discussion of this result mirrors that of squeezing.

	Similarly, we show in App. \ref{App: wigner} that the Wigner function, when computed using the APP, is nonnegative, which, together with the fact that there is no squeezing, implies that the state is either coherent or a mixed state \cite{Hudson1974}.

	Combining these findings, we draw a physical picture of the APP representation: By approximating the driving field as a statistical mixture, the nonclassicality of the driving field is neglected. This maps onto the generated field state, which also lacks these quantum optical features (quadrature squeezing below vacuum fluctuations, sub-Poissonian photon statistics, and negativity of the Wigner Function), and points to the generated field also being approximated as a statistical mixture of classical fields, as is consistent with the expectations discussed in Sec. \ref{Sec: APP}.
	
	Hence, by assuming that driving HHG with a coherent state generates a coherent state and by using the APP representation, the quantum optical features are neglected, and one can therefore not look for such quantum optical features in the generated light (or driving field) using these two approximations.

	\subsection{Example: Driving field} \label{Sec: driving}

	To see an example of the statements presented in Sec. \ref{Sec: General}, and to obtain a quantitative estimate of the error introduced by the APP, we now apply the APP representation to the driving field and compare the predicted expressions for the photonic observables against exact expressions.
	This case was already considered in Ref. \cite{Gothelf2025}, but is stated more generally in this work. We shall denote the driving field with the density operator $\hat{\rho}_L$, and the APP representation as 
	\begin{align}
		\hat{\rho}_L^{(\text{APP})} = \int d^2 \alpha Q(\alpha) \dyad{\alpha}{\alpha}.
	\end{align}

	Using the optical equivalence theorem for anti-normally ordered operators \cite{Walls2008}, we find expressions for the exact expectation values
	\begin{subequations}
		\label{Eq: driving_exact_expectations}
		\begin{align}
			\mean{\hat{a} \hat{a}^\dag} &= \int d^2\alpha Q(\alpha) \abs{\alpha}^2 \\
			\mean{\hat{a}^2} &= \int d^2\alpha Q(\alpha) \alpha^2 \\
			\mean{\hat{a}} &= \int d^2\alpha Q(\alpha) \alpha \\
			\mean{\hat{a} \hat{a} \hat{a}^\dag \hat{a}^\dag} &= \int d^2\alpha Q(\alpha) \abs{\alpha}^4,
		\end{align}\end{subequations} 
	and $\mean{\hat{a}^{\dag}}=\mean{\hat{a}}^*$ and $\mean{\hat{a}^{\dag 2}}=\mean{\hat{a}^2}^*$.
	
	Meanwhile, to find expectation values of normally ordered operators using the APP representation, we evaluate the traces
	\begin{subequations}
		\label{Eq: driving_APP_expectations}
		\begin{align}
			\mean{\hat{a}^\dag \hat{a}}^{(\text{APP})} = \text{Tr}\bigg[\hat{a}^\dag \hat{a} \hat{\rho}_L^{(\text{APP})}\bigg] &= \int d^2\alpha Q(\alpha) \abs{\alpha}^2 \\
			\mean{\hat{a}^2}^{(\text{APP})} =\text{Tr}\bigg[\hat{a}^2 \hat{\rho}_L^{(\text{APP})}\bigg] &=  \int d^2\alpha Q(\alpha) \alpha^2 \\
			\mean{\hat{a}}^{(\text{APP})} = \text{Tr}\bigg[\hat{a} \hat{\rho}_L^{(\text{APP})}\bigg]  &=  \int d^2\alpha Q(\alpha) \alpha \\
			\mean{\hat{a}^\dag \hat{a}^\dag \hat{a} \hat{a}}^{(\text{APP})} = \text{Tr}\bigg[\hat{a}^\dag \hat{a}^\dag \hat{a} \hat{a} \hat{\rho}_L^{(\text{APP})}\bigg] &= \int d^2\alpha Q(\alpha) \abs{\alpha}^4,
		\end{align}
	\end{subequations}
	and $\mean{\hat{a}^{\dag 2}}^{(\text{APP})} \!\!= \!\! \big[\mean{\hat{a}^2}^{(\text{APP})}\big]^*$ and $\mean{\hat{a}^{\dag}}^{(\text{APP})} \!\!= \!\! \big[\mean{\hat{a}}^{(\text{APP})}\big]^*$. We note that Eq. (\ref{Eq: driving_APP_expectations}) is equivalent to Eq. (\ref{Eq: general_APP_a_values}) using the function $\gamma_n^\alpha=\alpha$, and the statements about the quantum optical observables from Sec. \ref{Sec: General} thus apply to the driving field as well.
	
	Comparing Eqs. (\ref{Eq: driving_exact_expectations}) and (\ref{Eq: driving_APP_expectations}) we find that 
	\begin{subequations}
		\begin{align}
			\mean{\hat{a}^\dag \hat{a}}^{(\text{APP})}&=\mean{\hat{a} \hat{a}^\dag}=\mean{\hat{a}^\dag \hat{a}}+1 \\
			\mean{\hat{a}^2}^{(\text{APP})}&=\mean{\hat{a}^2} \\
			\mean{\hat{a}}^{(\text{APP})}&=\mean{\hat{a}} \\
			\mean{\hat{a}^\dag \hat{a}^\dag \hat{a} \hat{a}}^{(\text{APP})} &= \mean{\hat{a} \hat{a} \hat{a}^\dag \hat{a}^\dag}.
		\end{align}
		\label{Eq: driving_APP_vs_exact}
	\end{subequations}

	First, we see that the photon number $\mean{\hat{a}^\dag \hat{a}}^{(\text{APP})}$ differs from the exact result by a constant value of 1. For intense fields, the mean photon number is very large, and the error introduced by the APP is therefore small.

	However, for squeezing we see by inserting Eq. (\ref{Eq: driving_APP_vs_exact}) into Eq. (\ref{Eq: squeezing_operator}) that
	\begin{align}
		\mean{\Delta \hat{X}^2(\theta)}^{(\text{APP})} = \mean{\Delta \hat{X}^2(\theta)}+\frac{1}{2},
	\end{align}
	which has multiple consequences: In the APP representation, the quadrature variance is bounded from below by $1/2$, independent of the quantum nature of the field. Thus, quadrature squeezing cannot be found in the driving field when using the APP representation, as is consistent with Sec. \ref{Sec: General}. Furthermore, as a coherent state has  $\mean{\Delta \hat{X}^2(\theta)}=1/4$ for all $\theta$, we find that the APP representation approximates a coherent state as a state that is not of minimum quadrature variance, i.e., a state that is not a coherent state.

	Likewise, the second-order coherence function is found by inserting Eq. (\ref{Eq: driving_APP_vs_exact}) into Eq. (\ref{Eq: g2_form}).
	\begin{align}
		&g^{(2)\:(\text{APP})}(0) = \frac{\mean{\hat{a}^\dag \hat{a}^\dag \hat{a}\hat{a}}^{(\text{APP})}}{[\mean{\hat{a}^\dag\hat{a}}^{(\text{APP})}]^2} = \frac{\mean{\hat{a} \hat{a} \hat{a}^\dag\hat{a}^\dag}}{\mean{\hat{a}\hat{a}^\dag}^2}\notag \\
		=\:& \frac{\mean{\hat{a}^\dag\hat{a}^\dag\hat{a}\hat{a} }+4\mean{\hat{a}^\dag\hat{a}}+2}{\big(\mean{\hat{a}^\dag\hat{a}}+1\big)^2}
		\notag \\
		=\:& \frac{\mean{\hat{a}^\dag\hat{a}^\dag\hat{a}\hat{a} }}{\mean{\hat{a}^\dag\hat{a}}^2} \frac{\mean{\hat{a}^\dag\hat{a}}^2}{\big(\mean{\hat{a}^\dag\hat{a}}+1\big)^2} + \frac{4\mean{\hat{a}^\dag \hat{a}}+2}{\big(\mean{\hat{a}^\dag \hat{a}}+1\big)^2}
		\notag \\
		=\:&g^{(2)}(0) \frac{\mean{\hat{a}^\dag \hat{a}}^2}{\big(\mean{\hat{a}^\dag \hat{a}}+1\big)^2} + \frac{4\mean{\hat{a}^\dag \hat{a}}+2}{\big(\mean{\hat{a}^\dag \hat{a}}+1\big)^2},
	\end{align}
	where we have used the standard commutation relation of $\hat{a}$ and $\hat{a}^\dag$ to rewrite the expression.
	
	The error in this expression goes to zero as the driving field becomes strong, since $\frac{\mean{\hat{a}^\dag \hat{a}}^2}{(\mean{\hat{a}^\dag \hat{a}}+1)^2}\rightarrow1$ and $\frac{4\mean{\hat{a}^\dag \hat{a}}+2}{(\mean{\hat{a}^\dag \hat{a}}+1)^2}\rightarrow 0$ as $\mean{a^\dag a}\rightarrow\infty$. Specifically, for a Fock state we find that $g^{(2)\:(\text{APP})}(0)=1+\frac{1}{\mean{a^\dag a}+1}>1$, as is also predicted by Sec. \ref{Sec: General}. Comparing this to the exact result $g^{(2)}(0)=1-1/\mean{a^\dag a}$, we see that both expressions go to 1 for $\mean{a^\dag a}\rightarrow\infty$. Hence the absolute error goes to zero, but as $g^{(2)\:(\text{APP})}(0)$ approaches 1 from above while $g^{(2)}(0)$ approaches 1 from below, we see for all Fock states that though the error becomes small, the physical nature of the Fock state is mischaracterized as having super-Poissonian photon statistics, rather than the exact sub-Poissonian statistics. This discussion is consistent with the discussion of the Mandel parameter in the appendix of Ref. \cite{Gothelf2025}, which further illustrates the mischaracterization of the photon statistics for a Fock state driving field when using the APP.

	Thus, the intensity of the driving field in the APP representation contains a small error compared to exact predictions, but the quantum optical characteristics, i.e., the degree of squeezing and photon statistics, are not accurately captured.

	\subsection{Example: HHG from a one-band model} \label{Sec: one-band}
	\subsubsection{Exact results}\label{Sec: one-band-exact}

	In Sec. \ref{Sec: General}, we showed that the APP representation does not capture the nonclassical properties of light, if present, and in Sec. \ref{Sec: driving}, we considered the quantitative error this introduces to the driving field. However, as argued in Ref. \cite{Stammer2025b}, this does not immediately imply that the error in the observables of the generated field is of significant magnitude.

	For this reason, we now consider an example of the field generated from HHG in a one-dimensional single-band model of a solid driven by coherent light as considered in Ref. \cite{Gothelf2025}. As shown in Refs. \cite{Lange2024a,Gothelf2025} there are no dipole fluctuations, and Eq. (\ref{Eq: coherent_solution}) is therefore exact, i.e., the exact quantum state of the emitted light is a multi-mode coherent state.
	Furthermore, the analytical solutions for the electronic dynamics are obtainable, which allows us to obtain analytical expressions for the quantum optical observables of the emitted field.

	The Hamiltonian for the one-dimensional, one-band model driven by a classical field with vector potential $A^\alpha_{\text{cl}}(t)$ (polarized along the direction of the chain and with $\alpha$ again being the corresponding coherent state parameter) is, in crystal momentum space, given as \cite{Gothelf2025}
	\begin{align}
		\hat{H}^\alpha_{\text{sc}}(t) = \sum_{q,\mu} \mathcal{E}\big[q+A^\alpha_{\text{cl}}(t)\big]\hat{c}_{q,\mu}^\dag \hat{c}_{q,\mu},
		\label{Eq: intraband_Hamilton}
	\end{align}
	with $\hat{c}_{q,\mu}$ being the annihilation operator of the electron Bloch state with crystal momentum $q$ in the direction of the chain and spin $\mu$, and with 
	\begin{align}
		\mathcal{E}(q) = \sum_l b_l \cos(alq), \label{Eq: band-structure}
	\end{align}
	being the dispersion relation of the material with lattice constant $a$ and the $l$'th Fourier coefficient $b_l$.

	In this system, the current operator is given as \cite{Gothelf2025,Andersen2024,Lange2024a}
	\begin{align}
		\hat{j}^\alpha_{\text{sc}}(t) = \sum_{q,\mu} \frac{\partial \mathcal{E}\big[q+A^\alpha_{\text{cl}}(t)\big]}{\partial q}\hat{c}_{q,\mu}^\dag \hat{c}_{q,\mu}.
		\label{Eq: intraband_current}
	\end{align}
	From Eqs. (\ref{Eq: intraband_Hamilton}) and (\ref{Eq: intraband_current}), we see that $[\hat{H}^\alpha_{\text{sc}}(t),\hat{j}^\alpha_{\text{sc}}(t)] = [\hat{\mathcal{U}}_{\text{sc}}^\alpha(t),\hat{j}^\alpha_{\text{sc}}(t)] = 0$, allowing for a set of semiclassical solutions $\{\ket{\phi^\alpha_n(t)}\}$, i.e., $i\partial_t\ket{\phi^\alpha_n(t)} = \hat{H}^\alpha_{\text{sc}}(t)\ket{\phi^\alpha_n(t)}$ that are eigenstates of the current operator, i.e., $\hat{j}^\alpha_{\text{sc}}(t)  \ket{\phi^\alpha_n(t)}\propto\ket{\phi^\alpha_n(t)}$, which means that $\mel{\phi_m^\alpha(t)}{\hat{j}^\alpha_{\text{sc}}(t)}{\phi_n^\alpha(t)}\propto\delta_{m,n}$.

	Therefore, the solution is analytically given by Eq. (\ref{Eq: coherent_solution}), i.e., a multi-mode coherent state. We now go further by assuming a time-periodic pulse, which allows the current to be computed as \cite{Gothelf2025}
	\begin{align}
		j^\alpha_{i,i}(t) = -2a\sum_l C_l \sin\big[\tilde{g}_0 l \abs{\alpha}\sin(\omega_L t -\varphi_\alpha)\big],
		\label{Eq: intraband_current_element}
	\end{align}
	with $C_l = l b_l \sum_q \cos(alq)$ being a crystal specific geometric constant, $\tilde{g}_0 = 2ag_0/\sqrt{\omega_L}$ being a lattice-modified coupling constant and $\varphi_\alpha$ being the phase of $\alpha$.

	To compute the Fourier transform of Eq. (\ref{Eq: intraband_current_element}), we apply the Jacobi-Anger expansion and obtain
	\begin{align}
		\int_{-\infty}^\infty dt j^\alpha_{i,i}(t) e^{i\omega t} = -4i\pi a\sum_l C_l \sum_{n=1,3,5,\dots}^\infty J_n(l\tilde{g}_0 \abs{\alpha}) \notag \\
		\times \big[e^{-in\phi}\delta(\omega+n\omega_L)-e^{in\phi}\delta(\omega-n\omega_L)\big],
		\label{Eq: current_fourier_transform}
	\end{align}
	where $J_n$ denotes the $n$'th order Bessel function of the first kind. Substituting Eq. (\ref{Eq: current_fourier_transform}) into Eq. (\ref{Eq: gamma_expression}) and only considering positive frequencies, we obtain the analytical expression for the coherent amplitude of the odd harmonics
	\begin{align}
		\gamma^\alpha_{n} = G_n\sum_l C_l J_n(l\tilde{g}_0 \abs{\alpha})e^{in\varphi_\alpha},
		\label{Eq: gamma_oneband}
	\end{align}
	where $n$ is an odd integer and $G_n = \frac{4\pi a g_0}{\sqrt{\omega_n}} \delta(\omega_n-n\omega_L)$ is a coefficient determining the amplitude of the coherent state in harmonic $n$.

	We note that the presence of a delta function in $G_n$ is due to the infinite nature of the time-periodic pulse. Renormalization is therefore done for numerical purposes by matching the amplitude to a simulation of a pulse with finite duration \cite{Gothelf2025}.

	Equation (\ref{Eq: gamma_oneband}) is exact and can be used to compute observables of interest exactly. Specifically, the spectrum is
	\begin{align}
		S(\omega_n) \propto \mean{\hat{a}_n^\dag \hat{a}_n}=|\gamma^\alpha_n|^2,
	\end{align}
	which can be evaluated using Eq. (\ref{Eq: gamma_oneband}), and where the sum over $\sigma$ is neglected, as only the polarization direction along the lattice contributes \cite{Gothelf2025}.
	
	For the consideration of quantum optical observables in the presented one-band model, we choose to consider only the quadrature squeezing for simplicity. The exact value of the quadrature variance is
	\begin{align}
		\mean{\Delta \hat{X}_n^2(\theta)} = \frac{1}{4},
		\label{Eq: squeezing_exact}
	\end{align}
	as the state is a coherent state.
	
	To be able to compare the exact expressions of this section against the expressions using the APP representation in Sec. \ref{Sec: oneband_APP_predict} below, we expand Eq. (\ref{Eq: gamma_oneband}) in terms of the driving field amplitude, $\abs{\alpha}$. Doing so yields the expression
	\begin{align}
		\gamma_n^\alpha = G_n\sum_{m=0}^\infty D^{(n)}_m \abs{\alpha}^{2m+n} e^{in\varphi_\alpha},
		\label{Eq: gamm_taylor_expansion}
	\end{align}
	with $D^{(n)}_m = \sum_l (-1)^m C_l \Big(\frac{\tilde{g}_0}{2}\Big)^{2m+n}\frac{l^{2m+n}}{m!(m+n)!}$ being expansion coefficients, stemming from the power-series expansions of the Bessel functions. From this expression, we find that
	\begin{align}
		\mean{\hat{a}_n^\dag \hat{a}_n} &= G_n^2 \sum_{m_1=0}^\infty \sum_{m_2=0}^\infty D^{(n)}_{m_1} D^{(n)}_{m_2} \abs{\alpha}^{2m_1+2m_2+2n} ,
		\label{Eq: a_dag_a_oneband_exact}
	\end{align}
	with similar expressions for $\mean{\hat{a}_n^2}$ and $\mean{\hat{a}_n}$ given in App. \ref{App: details_oneband}.

	\subsubsection{APP predictions}\label{Sec: oneband_APP_predict}
	We now calculate the same observables as considered in Sec. \ref{Sec: one-band-exact} above for the coherently driven one-band model using the APP representation and compare them to the exact calculations.

	In Eq. (\ref{Eq: a_dag_a_oneband_exact}), the exact expression of the expectation value $\mean{\hat{a}_n^\dag\hat{a}_n}=\mel{\gamma_n^\alpha}{\hat{a}_n^\dag\hat{a}_n}{\gamma_n^\alpha}$ is given for a coherent driving field $\alpha$. To obtain the APP expression for this expectation value, we apply Eq. (\ref{Eq: APP_expectation_value}), i.e., we integrate the exact expression for a coherently driven system [Eq. (\ref{Eq: a_dag_a_oneband_exact})] with the Husimi function of the driving field, which for a coherent state $\ket{\alpha}$ is given as
	\begin{align}
		Q(\beta) = \frac{1}{\pi} e^{-\abs{\beta-\alpha}^2},
		\label{Eq: Coherent_Q}
	\end{align}
	where $\beta$ is the phase-space coordinate for integration. 
	
	Doing so, we find that
	\begin{align}
		&\mean{\hat{a}_n^\dag \hat{a}_n}^{(\text{APP})} = G_n^2 \sum_{m_1=0}^\infty \sum_{m_2=0}^\infty D^{(n)}_{m_1} D^{(n)}_{m_2} \notag\\ 
		&\quad\times \sum_{q=0}^{m_1+m_2+n}\binom{m_1+m_2+n}{q}\frac{(m_1+m_2+n)!}{q!}\abs{\alpha}^{2q}, 
		\label{Eq: a_dag_a_oneband_APP}
	\end{align}
	with details on the calculations and similar expressions for $\mean{\hat{a}_n^2}^{(\text{APP})}$ and $\mean{\hat{a}_n}^{(\text{APP})}$ given in App. \ref{App: details_oneband}.
	
	Comparing the APP expression for the spectrum in Eq. (\ref{Eq: a_dag_a_oneband_APP}) to the exact expression in Eq. (\ref{Eq: a_dag_a_oneband_exact}) by terms in the sum over $m_1$ and $m_2$, we see that the highest-order term in the sum over $q$ in Eq. (\ref{Eq: a_dag_a_oneband_APP}) matches Eq. (\ref{Eq: a_dag_a_oneband_exact}). Thus, the lower-order terms in Eq. (\ref{Eq: a_dag_a_oneband_APP}) account for the error introduced by the APP representation.

	For the spectrum, the termwise relative error in the sum over $m_1$ and $m_2$ is found by subtracting the terms in Eqs. (\ref{Eq: a_dag_a_oneband_exact}) and (\ref{Eq: a_dag_a_oneband_APP}) and dividing by the term in the exact expression [Eq. (\ref{Eq: a_dag_a_oneband_exact})], i.e., 
	\begin{align}
		&\frac{\sum_{q=0}^{m_1+m_2+n}\allowbreak\binom{m_1+m_2+n}{q}\frac{(m_1+m_2+n)!}{q!}\abs{\alpha}^{2q} - \abs{\alpha}^{2m_1+2m_2+2n}}{\abs{\alpha}^{2m_1+2m_2+2n}}
		\notag
		\\
		&=\!\!\!\!\!\!\!\!  \sum_{q=0}^{m_1+m_2+n-1}\!\!\!\! \binom{m_1+m_2+n}{q}\frac{(m_1+m_2+n)!}{q!}\abs{\alpha}^{2q-2(m_1+m_2+n)}
		\notag 
		\\
		&=\mathcal{O}(1/\abs{\alpha}^2).
	\end{align}
	 We may therefore expect the relative error on the spectrum to go approximately as $\mathcal{O}(1/\abs{\alpha}^2)$, and since $\abs{\alpha}^2$ is proportional to the driving field intensity, we see that the relative error on the spectrum vanishes for strong driving fields, which is consistent with results presented in Refs. \cite{Wang2025,Gonzalez-Monge2025,Gothelf2025}, as previously stated.

	For the quadrature squeezing, we saw for the exact result, since the emitted state is coherent, that
	\begin{align}
		\mean{\hat{a}_n^\dag \hat{a}_n}-\mean{\hat{a}_n^\dag}\mean{ \hat{a}_n}&=0 \notag \\
		\mean{\hat{a}_n^2}-\mean{ \hat{a}_n}^2&=0,
		\label{Eq: a_diff_exact}
	\end{align}
	and by inserting Eq. (\ref{Eq: a_diff_exact}) into Eq. (\ref{Eq: squeezing_operator}), we recover the result in Eq. (\ref{Eq: squeezing_exact}).
	
	Using the APP expressions, which are written in detail in App. \ref{App: details_oneband}, we find that
	\begin{align}
		\mean{\hat{a}_n^\dag \hat{a}_n}^{(\text{APP})}-\mean{\hat{a}_n^\dag}\mean{ \hat{a}_n}^{(\text{APP})}&\neq0 \notag \\
		\mean{\hat{a}_n^2}^{(\text{APP})}-\Big[\mean{ \hat{a}_n}^{(\text{APP})}\Big]^2&\neq0,
		\label{Eq: a_diff_APP}
	\end{align}
	which by insertion into Eq. (\ref{Eq: squeezing_operator}) yields $\mean{\Delta \hat{X}_n^2(\theta)}^{(\text{APP})} \neq \frac{1}{4}$. 
	Comparing Eqs. (\ref{Eq: a_diff_exact}) and (\ref{Eq: a_diff_APP}), we see that the exact expressions are vanishing, whereas the APP expressions are not. The error introduced by the APP is thus significant in this instance. 
	
	A similar analysis could be considered for $g^{(2)}(0)$ by considering higher moments of $\hat{a}_n$ and $\hat{a}_n^\dag$. We expect a small positive error, though the scaling of the error with system size as seen for the squeezing is not expected for $g^{(2)}(0)$ due to the division by $\mean{\hat{a}_n^\dag\hat{a}_n}^2$ in Eq. (\ref{Eq: g2_form}), which acts a normalization. This expectation is consistent with the considerations for the driving field in Sec. \ref{Sec: driving}.

	\subsubsection{Numerical results}
	
	To estimate the error in the quadrature variance quantitatively, we consider the highest-order nonvanishing terms in the quadrature squeezing calculated using the APP expressions. For the full expressions, see Eq. (\ref{Eq: a_dag_a_oneband_APP}) and App. \ref{App: details_oneband}. The highest-order nonvanishing term inside the sum over $m_1$ and $m_2$ in the expansion of Eq. (\ref{Eq: a_diff_APP}) is of order $\abs{\alpha}^{2(m_1+m_2+n-1)}$, and Eq. (\ref{Eq: a_diff_APP}) is determined to the highest order as 
	\begin{align}
		&\mean{\hat{a}_n^\dag\hat{a}_n}^{(\text{APP})}-\mean{\hat{a}_n^\dag}^{(\text{APP})}\mean{\hat{a}_n}^{(\text{APP})} 
		\notag \\
		&\qquad\approx G_n^2 \sum_{m_1}^\infty\sum_{m_2}^\infty D_{m_1}^{(n)}D_{m_2}^{(n)} \abs{\alpha}^{2(m_1+m_2+n-1)}
		\notag \\
		&\qquad\quad\times\big(n^2+2m_1m_2+nm_2+nm_2\big)
		\label{Eq: squeez_first_term_highest_error_main}
	\end{align}
	and
	\begin{align}
		&\mean{\hat{a}_n^2}^{(\text{APP})}-\big[\mean{\hat{a}_n}^{(\text{APP})} \big]^2
		\notag \\
		&\qquad\approx G_n^2 e^{2in\varphi_\alpha} \sum_{m_1}^\infty\sum_{m_2}^\infty D_{m_1}^{(n)}D_{m_2}^{(n)} \abs{\alpha}^{2(m_1+m_2+n-1)}
		\notag \\
		&\qquad\quad\times\big(2m_1m_2+nm_2+nm_2\big).
		\label{Eq: squeez_second_term_highest_error_main}
	\end{align}
	
	As explained in detail in App. \ref{App: details_oneband}, Eqs. (\ref{Eq: squeez_first_term_highest_error_main}) and (\ref{Eq: squeez_second_term_highest_error_main}) have closed-form expressions in terms of Bessel functions. Hence, numerical evaluation of the highest-order term for the quadrature variance in harmonic $n$ is possible. Using that the exact value of the squeezing is $1/4$ [Eq. (\ref{Eq: squeezing_exact})], we compute the error from this highest-order expansion as
	\begin{align}
		&\mean{\Delta \hat{X}_n^2(\theta)}^{(\text{APP})} - \mean{\Delta \hat{X}_n^2(\theta)} \notag \\
		&= \frac{1}{4}\bigg\{2\Big(\mean{\hat{a}_n^\dag\hat{a}_n}^{(\text{APP})}-\mean{\hat{a}_n^\dag}^{(\text{APP})}\mean{\hat{a}_n}^{(\text{APP})}\Big)
		\notag \\
		&\quad + \Big[e^{-2i\theta}\Big(\mean{\hat{a}_n^2}^{(\text{APP})}-\mean{\hat{a}_n}^{(\text{APP})\:2}\Big) + c.c.\Big]\bigg\}.
		\label{Eq: error_in_squeezing}
	\end{align}
	Similarly, the next-highest-order terms in $\mean{\hat{a}_n^\dag\hat{a}_n}^{(\text{APP})}-\mean{\hat{a}_n^\dag}^{(\text{APP})}\mean{\hat{a}_n}^{(\text{APP})}$ and $\mean{\hat{a}_n^2}^{(\text{APP})}-\big[\mean{\hat{a}_n}^{(\text{APP})} \big]^2$ can also be expressed in terms of Bessel functions, which yields an estimate of the size of the terms below the highest-order terms.

	For the evaluation of Eq. (\ref{Eq: error_in_squeezing}), we use parameters similar to Ref. \cite{Gothelf2025}, letting $g_0=4\times10^{-8}$, $\omega_L= 0.005$. and $\mean{\hat{a}_{\vec{k}_L,\sigma_L}^\dag\hat{a}_{\vec{k}_L,\sigma_L}}=7.35\times10^{11}$ (corresponding to an intensity of $I=8.26\times10^{11}$ W/$\text{cm}^2$). The laser pulse used for renormalization of Eq. (\ref{Eq: gamma_oneband}) is modeled as a 20 laser-cycle pulse with a $\sin^2$ envelope. For the one-band model, we use the lattice constant $a=5.32$, and the Fourier coefficients of the dispersion in Eq. (\ref{Eq: band-structure}) are chosen to match a ZnO crystal along the $\Gamma$-$M$ direction \cite{Vampa2015a} such that $b_1 = -0.0814$, $b_2 = -0.0024$, $b_3 = -0.0048$, $b_4 = -0.0003$, and $b_5 = -0.0009$.
	Periodic boundary conditions are applied with $L$ lattice sites such that the discretization of the crystal momentum is $\Delta q = \frac{2\pi}{L a}$, and a half-filling arrangement is assumed, meaning that $L$ electrons are considered.

\begin{figure}
		\includegraphics[width=\linewidth]{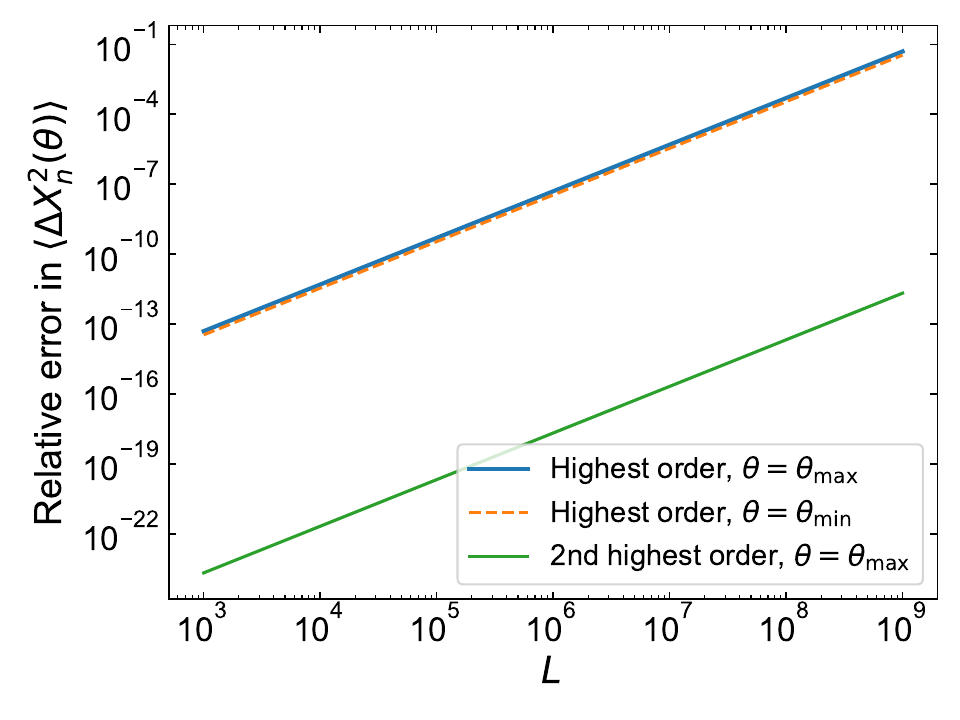}
		\caption{Relative error in quadrature variance in harmonic order $n=3$ introduced by the APP representation for a one-dimensional single-band solid as a function of the number of electrons in the material ($L$). Solid blue and dashed orange curves denote the error at the maximizing ($\theta_{\text{max}}$) and minimizing ($\theta_{\text{min}}$) angles, respectively, calculated from the highest-order terms in driving field intensity. The green curve is the second-highest-order term in the error at the maximizing angle $\theta_{\text{max}}$, and is seen not to contribute significantly.}
		\label{Fig: coherent_oneband_squeezing}
	\end{figure}

	In Fig. \ref{Fig: coherent_oneband_squeezing}, we plot, for harmonic order $n=3$, the maximum (blue curve) and minimum (orange curve) error in the quadrature variance over the angle $\theta$, i.e., $\max_{\theta\in [0,\pi)}\Big(\mean{\Delta \hat{X}_n^2(\theta)}^{(\text{APP})} - \mean{\Delta \hat{X}_n^2(\theta)}\Big)$ and likewise for the minimum, calculated using the highest-order expansion [Eqs. (\ref{Eq: squeez_first_term_highest_error_main})-(\ref{Eq: squeez_second_term_highest_error_main})], relative to the exact value of $1/4$. This relative error is plotted as a function of the number of electrons $L$ (which is the same as the number of lattice sites) to illustrate the scaling of the error with system size. 
	
	We note that both the minimum and maximum errors are positive, and that the maximum and minimum are not exactly equal. Thus, the quadrature variance for a coherent state computed using the APP representation is larger than the exact value and is dependent on the angle $\theta$. As the minimum of the error in quadrature variance is positive, we also see that the state is antisqueezed, as is consistent with Sec. \ref{Sec: General}, since the quadrature variance calculated using the APP is larger for all angles $\theta$ compared with the exact value.  
	
	Furthermore, we plot in Fig. \ref{Fig: coherent_oneband_squeezing} the maximum error in the quadrature variance (green curve) calculated using the second-highest-order expansion. We see that the error stemming from the second-highest-order is several orders of magnitude smaller than the error from the highest-order term for the chosen parameters. Thus, we conclude that the highest-order terms capture the error.

	From Fig. \ref{Fig: coherent_oneband_squeezing}, we see a scaling of the error in quadrature squeezing with the system size that is quadratic (notice the log-log scale in Fig. \ref{Fig: coherent_oneband_squeezing}).
	As such, even though the error is small for small systems, the error can become large if the system is large. 
	
	Indeed, this scaling is consistent with what one would expect from the expression for the quadrature variance [Eq. (\ref{Eq: squeezing_operator})]: If there is an error in squeezing and we amplify the coherent amplitudes by $\gamma_n^\alpha \mapsto K \gamma_n^\alpha$ for some real number $K$, then there is a gain in the squeezing error by $K^2$. The same scaling will therefore also be seen from a longer pulse duration, as the harmonic peaks become sharper. 
	
	Even under the arguments for introducing the APP presented in Ref. \cite{Stammer2025b,Rivera-Dean2025a} of taking $g_0\rightarrow 0$ while keeping $g_0\abs{\alpha}$ and $N g_0$ constant, with $N$ being the number of emitters, we see a scaling in the error with pulse duration and emitter density.
	
	For systems where the exact quadrature variance is not $1/4$, such as the atomic and solid state systems considered in Refs. \cite{Gorlach2020,Lange2024a,Lange2025b,Lange2025c}, the exact squeezing is similarly expected to scale with the coherent amplitudes \cite{Lange2025c}. The exact value and the error would therefore become larger in tandem as the coherent amplitudes become larger. The significance of the error for such systems would then depend on the relative scaling between the exact value and error, and whether the error and exact values are of the same orders of magnitude. Further analysis of such systems is left for future work, though we know that squeezing below quadrature fluctuations, as found in these works, cannot be captured using the APP representation, as follows from Sec. \ref{Sec: General}. 
	
	We have thus seen that the APP representation introduces a relative error in the quadrature variance, which is small for a few-emitter system. However, this error is amplified in larger electronic systems and by using a driving pulse of longer duration. For a coherent state, which should be a minimum quadrature variance state, where the quadrature variance is uniform over the angle $\theta$, the APP predicts a variance that is larger than the exact result, and depends on $\theta$, i.e., antisqueezing is falsely predicted. For more complex systems and driving fields, the error introduced by the APP is not controlled. 
	Hence, to use the APP to calculate quantum observables, such as squeezing in the quadrature variance, the error introduced by the APP in the specific system should be assessed before physical meaning is attributed to the results. 
	
	\section{Conclusion and outlook}\label{Sec: conclusion}
	
	In this paper, we investigated the accuracy of the approximative phase-space representation for driving HHG with a general quantum state of light.

	We have described the APP representation and applied it to compute relevant expectation values. We analyzed the error introduced by the APP with regard to the observables of interest: The HHG spectrum, quadrature squeezing, and photon statistics. 
	
	We have shown that if one assumes that the generating electronic medium has no dipole correlations, then quadrature squeezing below vacuum fluctuations and sub-Poissonian photon statistics cannot be captured by the APP representation, independent of the quantum nature of the driving field.

	Examining the accuracy of the APP representation for the driving field, we find that the relative error on the intensity is negligible at high intensity, while quadrature squeezing is mischaracterized, and the photon statistics of Fock states are not captured.

	Lastly, we also tested the APP representation on the HHG field emitted from a one-band model driven by coherent light.
	From this, we found that the relative error on the spectrum is diminishing for higher intensity driving fields, while for squeezing, the absolute error can become significant, as it scales with pulse duration and emitter density, though the error is small in small systems.

	In conclusion, by introducing the APP representation, the driving field is approximated as a statistical mixture of classical fields, and the field generated in the HHG process inherits this characteristic, i.e., any present nonclassical features are neglected.

	Furthermore, an understanding of why an observable may be captured in the APP or not can be drawn. For classical measures like the spectrum, the relative error diminishes with driving field intensity. However, for relative measures such as fluctuations (quadrature squeezing and photon statistics), in which expectation values are compared by, e.g., subtraction, and which are used to characterize nonclassical states of light, the error can become significant. For precise statements on the accuracy of other observables, further analysis is needed.

	Therefore, to study these observables, either the system-specific error has to be quantified and accounted for, or other methods than the APP need to be utilized. Alternative methods have already been presented: In Ref. \cite{Wang2025}, a generalized von Neumann lattice basis was used for the coherent state expansion. In Ref. \cite{EvenTzur2024}, a perturbative approach was employed to consider the mixture of a strong coherent and a perturbative squeezed field. In Ref. \cite{Gonzalez-Monge2025}, a correction term was considered in the continuum limit, and in Ref. \cite{Mao2025}, a Feynman path integral approach was used to consider the joint electron-photon spectrum from above-threshold ionization. 
	
	\begin{acknowledgements}
		This work is supported by the Independent Research Fund Denmark (Technology and Production Sciences 10.46540/4286-00053B) and the Novo Nordisk Foundation Project Grants in the Natural and Technical Sciences (0094623).
	\end{acknowledgements} 
	
	\appendix

	\section{Nonnegativity of the Wigner function}\label{App: wigner}
	In this appendix, we support the discussion in Sec. \ref{Sec: General} by showing that the Wigner function is nonnegative when calculated using the APP in the absence of dipole correlations.
	
	The Wigner function for a mode $n$ is the quasi-probability distribution for a photonic state, defined as the two-dimensional Fourier transform of the expectation value of the displacement operator \cite{Gerry2004}
	\begin{align}
		W_n(\alpha) = \frac{1}{\pi^2}\int d^2\lambda e^{\lambda^*\alpha - \lambda \alpha^*} \mean{\hat{D}_n(\lambda)},
		\label{Eq: Wigner_definition}
	\end{align}
	where $\hat{D}_n(\lambda)=e^{\lambda \hat{a}_n^\dag - \lambda^* \hat{a}_n^\dag}$ is the displacement operator in mode $n$ with complex displacement $\lambda$. The photonic state is specified by the state used to compute the expectation value of the displacement operator.

	Negativity in the Wigner function signifies a nonclassical state \cite{Gerry2004}. Furthermore, Hudson's theorem \cite{Hudson1974} says for pure states that the state is a displaced squeezed state if and only if its Wigner function is nonnegative. 
	
	In the APP representation, using the same assumptions as in Sec. \ref{Sec: General}, we find that 
	\begin{align}
		\mean{\hat{D}_n(\lambda)}^{(\text{APP})}
		=&\int d^2\beta Q(\beta) \mel{\gamma_n^\beta}{\hat{D}_n(\lambda)}{\gamma_n^\beta}
		\notag \\
		=&
		\int d^2\beta Q(\beta)e^{i\text{Im}(\gamma_n^{\beta*}\lambda)}\inner{\gamma_n^\beta}{\gamma_n^\beta+\lambda}
		\notag \\
		=&\int d^2\beta Q(\beta)e^{2i\text{Im}(\gamma_n^{\beta*}\lambda)}e^{-\frac{1}{2}\abs{\lambda}^2},
	\end{align}
	which is inserted into Eq. (\ref{Eq: Wigner_definition}) to get
	\begin{align}
		W_n(\alpha) =& \frac{1}{\pi^2}\int d^2\lambda e^{\lambda^*\alpha - \lambda \alpha^*} \int d^2\beta Q(\beta) e^{2i\text{Im}(\gamma_n^{\beta*}\lambda)}e^{-\frac{1}{2}\abs{\lambda}^2}
		\notag \\
		=& \frac{1}{\pi^2} \int d^2\beta Q(\beta) \int d^2\lambda e^{\lambda^*(\alpha-\gamma_n^{\beta*}) - \lambda (\alpha-\gamma_n^{\beta*})^*}e^{-\frac{1}{2}\abs{\lambda}^2}.
	\end{align}
	The inner integral is of the type (letting $\lambda=x+iy$ and $\eta=a+ib$)
	\begin{align}
		& \int d^2\lambda e^{\lambda^*\eta - \lambda \eta^*}e^{-\frac{1}{2}\abs{\lambda}^2}
		\notag \\
		&= \int_{-\infty}^\infty dx e^{2ibx}e^{-\frac{1}{2}x^2}\int_{-\infty}^\infty dy e^{-2iay}e^{-\frac{1}{2}y^2}
		\notag \\
		&=2\pi e^{-\frac{1}{2}(-2b)^2} e^{-\frac{1}{2}(2a)^2}
		\notag \\
		&=2\pi e^{-2\abs{\eta}^2} ,
	\end{align}
	where we have used the fact that the Fourier transform of a Gaussian is also a Gaussian. Setting $\eta= \alpha-\gamma_n^{\beta*}$ we find that
	\begin{align}
		W_n(\alpha) = \frac{2}{\pi} \int d^2\beta Q(\beta) e^{-2\abs{\alpha-\gamma_n^{\beta*}}^2} \geq 0,
	\end{align}
	which is nonnegative, as both the Husimi function and the Gaussian function are nonnegative.
	
	We have thus shown that using the APP representation for a system that emits a coherent state when driven by a coherent state, negativity in the Wigner function cannot be found. 
	
	Furthermore, as also stated in Sec. \ref{Sec: General}, due to Hudson's theorem \cite{Hudson1974}, we may therefore conclude that the state in mode $n$ is either a displaced squeezed state or a mixed state. Since we showed in Sec. \ref{Sec: General} that it cannot be a squeezed state, we must conclude that it is either a coherent state or a mixed state.

	\section{Details on the APP considerations of a single-band model}\label{App: details_oneband}
	In this appendix we show the details of the derivation of the APP expressions for the expectation values $\mean{\hat{a}_n^\dag \hat{a}_n}^{(\text{APP})}$, $\mean{\hat{a}_n^2}^{(\text{APP})}$ and $\mean{\hat{a}_n}^{(\text{APP})}$, to aid results and discussion in Sec. \ref{Sec: oneband_APP_predict} of the main text.
	
	To find the APP expressions for the spectrum and squeezing, we begin by expressing the expectation values of the relevant observable in the Taylor expanded form using Eq. (\ref{Eq: gamm_taylor_expansion})
	\begin{subequations}
		\begin{align}
			\mean{\hat{a}_n^\dag \hat{a}_n} &= G_n^2 \sum_{m_1=0}^\infty \sum_{m_2=0}^\infty D^{(n)}_{m_1} D^{(n)}_{m_2} \abs{\alpha}^{2m_1+2m_2+2n} \label{Eq: a_dag_a_oneband_exact_appendix}\\
			\mean{\hat{a}_n^2} &= G_n^2 \sum_{m_1=0}^\infty \sum_{m_2=0}^\infty D^{(n)}_{m_1} D^{(n)}_{m_2} \abs{\alpha}^{2m_1+2m_2+2n} e^{2in\varphi_\alpha}  \label{Eq: a2_oneband_exact}\\
			\mean{\hat{a}_n} &= G_n^2 \sum_{m=0}^\infty D^{(n)}_{m} \abs{\alpha}^{2m+n} e^{in\varphi_\alpha}  \label{Eq: a_oneband_exact}.
		\end{align}
		\label{Eq: oneband_exact_appendix}
	\end{subequations}
	We integrate these expressions with the Husimi function of a coherent state [Eq. (\ref{Eq: Coherent_Q})]. First, we rewrite the Husimi function for a coherent state
	\begin{align}
		Q(\beta) &= \frac{1}{\pi}e^{-\abs{\alpha}^2-\abs{\beta}^2+2\text{Re}(\alpha^*\beta)} \notag \\
		&= \frac{1}{\pi}e^{-\abs{\alpha}^2-r^2+2r\abs{\alpha}\cos(\phi-\varphi_\alpha)},
	\end{align}
	where we have written $\beta=re^{i\phi}$ in polar form.

	We use the $2\pi$ periodicity of the integrand to see for an integer $m$ that
	\begin{align}
		&\int_0^{2\pi} d\phi e^{im\phi}e^{2r\abs{\alpha}\cos(\phi-\varphi_\alpha)} \notag \\
		&=e^{im\varphi_\alpha}\int_{-\pi}^{\pi} d\phi e^{im\phi}e^{2r\abs{\alpha}\cos(\phi)} \notag \\
		&=2e^{im\varphi_\alpha}\int_{0}^{\pi} d\phi \cos(im\phi)e^{2r\abs{\alpha}\cos(\phi)} \notag \\
		&= 2\pi e^{im\varphi_\alpha} I_{m}(2r\abs{\alpha}),
	\end{align}
	where $I_m$ denotes the $m$'th order modified Bessel function of the first kind. 
	
	From this we compute the APP expression for Eq. (\ref{Eq: a_dag_a_oneband_exact_appendix}) termwise in the sum over $m_1,m_2$
	\begin{align}
		&\int d^2\beta Q(\beta)\abs{\beta}^{2m_1+2m_2+2n} \notag\\
		&=	\frac{1}{\pi}e^{-\abs{\alpha}^2}\int_0^\infty dr e^{-r^2} r^{2(m_1+m_2+n)+1}\int_0^{2\pi}d\phi e^{2r\abs{\alpha}\cos(\phi-\varphi_\alpha)} \notag \\
		&= 2 e^{-\abs{\alpha}^2}\int_0^\infty dr e^{-r^2} r^{2(m_1+m_2+n)+1} I_0(2r\abs{\alpha}) \notag \\
		&= 2 e^{-\abs{\alpha}^2} \sum_{l=0}^\infty \int_0^\infty dr e^{-r^2} r^{2(m_1+m_2+n)+1} \frac{(r\abs{\alpha})^{2l}}{(l!)^2} \notag \\
		&=  e^{-\abs{\alpha}^2} \sum_{l=0}^\infty \abs{\alpha}^{2l} \frac{(l+m_1+m_2+n)!}{(l!)^2}.
		\label{Eq: term_APP_integral}
	\end{align}
	To proceed, we consider the function
	\begin{align}
		S_{i,j}(x) = x^{-i} \frac{d^j}{dx^j}(e^x x^{i+j}),
		\label{Eq: S_i,j}
	\end{align}
	for a real number $x$ and two natural numbers $i$ and $j$. Taylor expanding the exponential, we find that
	\begin{align}
		S_{i,j}(x) &= x^{-i} \frac{d^j}{dx^j}\sum_{l=0}^\infty \frac{x^{l+i+j}}{l!} \notag \\
		&= x^{-i} \sum_{l=0}^\infty \frac{(l+i+j)!}{(l+i)!}\frac{x^{l+i}}{l!} \notag \\
		&= \sum_{l=0}^\infty \frac{(l+i+j)!}{(l+i)!}\frac{x^{l}}{l!},
		\label{Eq: S_infite}
	\end{align}
	while applying Leibniz rule to Eq. (\ref{Eq: S_i,j}), we find
	\begin{align}
		S_{i,j}(x) &= x^{-i} \sum_{q=0}^j \binom{j}{q} \Big(\frac{d^q}{dx^q}e^x\Big) \Big(\frac{d^{j-q}}{dx^{j-q}}x^{i+j}\Big)\notag \\
		&= e^x \sum_{q=0}^j \binom{j}{q} \frac{(i+j)!}{(i+q)!} x^q,
		\label{Eq: S_finite}
	\end{align}
	which is a finite sum. We can thus convert infinite sums of the type in Eq. (\ref{Eq: S_infite}) to finite sums.

	This is applied to Eq. (\ref{Eq: term_APP_integral}), and we see
	\begin{align}
		&\int d^2\beta Q(\beta)\abs{\beta}^{2m_1+2m_2+2n} \notag\\
		&=  e^{-\abs{\alpha}^2} S_{0,m_1+m_2+n}(\abs{\alpha}^2) \notag \\
		&= \sum_{q=0}^{m_1+m_2+n}\binom{m_1+m_2+n}{q}\frac{(m_1+m_2+n)!}{q!}\abs{\alpha}^{2q}, 
		\label{Eq: term_APP_integral_adaga}
	\end{align}
	which is the termwise expression for $\mean{\hat{a}_n^\dag \hat{a}_n}^{(\text{APP})}$.

	Similarly, we can write the termwise expression for $\mean{\hat{a}_n^2}^{(\text{APP})}$ as
	\begin{align}
		&\int d^2\beta Q(\beta)\abs{\beta}^{2m_1+2m_2+2n}e^{2in\varphi_\alpha} \notag\\
		&= 2 e^{-\abs{\alpha}^2}e^{2in\varphi_\alpha}\int_0^\infty dr e^{-r^2} r^{2(m_1+m_2+n)+1} I_{2n}(2r\abs{\alpha}) \notag \\
		&=  e^{-\abs{\alpha}^2}e^{2in\varphi_\alpha} \abs{\alpha}^{2n} S_{2n,m_1+m_2}(\abs{\alpha}^2) \notag \\
		&=e^{2in\varphi_\alpha} \sum_{q=0}^{m_1+m_2}\binom{m_1+m_2}{q}\frac{(m_1+m_2+2n)!}{(q+2n)!}\abs{\alpha}^{2q+2n},
		\label{Eq: term_APP_integral_a^2}
	\end{align}
	and for $\mean{\hat{a}_n}^{(\text{APP})}$ as 
	\begin{align}
		&\int d^2\beta Q(\beta)\abs{\beta}^{2m+n}e^{in\varphi_\alpha} \notag\\
		&= 2 e^{-\abs{\alpha}^2}e^{in\varphi_\alpha}\int_0^\infty dr e^{-r^2} r^{2(m+n)+1} I_{n}(2r\abs{\alpha}) \notag \\
		&=  e^{-\abs{\alpha}^2}e^{in\varphi_\alpha} \abs{\alpha}^{n} S_{n,m}(\abs{\alpha}^2) \notag \\
		&=e^{in\varphi_\alpha} \sum_{q=0}^{m_1+m_2}\binom{m}{q}\frac{(m+n)!}{(q+n)!}\abs{\alpha}^{2q+n}.
		\label{Eq: term_APP_integral_a}
	\end{align}
	Putting Eqs. (\ref{Eq: term_APP_integral_adaga}), (\ref{Eq: term_APP_integral_a^2}) and (\ref{Eq: term_APP_integral_a}) into their respective sums, we achieve the following expressions for the APP expectation values
	\begin{subequations}
		\begin{align}
			&\mean{\hat{a}_n^\dag \hat{a}_n}^{(\text{APP})} = G_n^2 \sum_{m_1=0}^\infty \sum_{m_2=0}^\infty D^{(n)}_{m_1} D^{(n)}_{m_2} \notag\\ 
			&\quad\times \sum_{q=0}^{m_1+m_2+n}\binom{m_1+m_2+n}{q}\frac{(m_1+m_2+n)!}{q!}\abs{\alpha}^{2q} \label{Eq: a_dag_a_oneband_APP_appendix}\\
			&\mean{\hat{a}_n^2}^{(\text{APP})} = G_n^2 \sum_{m_1=0}^\infty \sum_{m_2=0}^\infty D^{(n)}_{m_1} D^{(n)}_{m_2} e^{2in\varphi_\alpha} \notag\\ 
			&\quad\times \sum_{q=0}^{m_1+m_2}\binom{m_1+m_2}{q}\frac{(m_1+m_2+2n)!}{(q+2n)!}\abs{\alpha}^{2q+2n} \label{Eq: a2_oneband_APP}\\
			&\mean{\hat{a}_n}^{(\text{APP})} = G_n \sum_{m=0}^\infty D^{(n)}_{m} e^{in\varphi_\alpha} 
			\sum_{q=0}^{m}\binom{m}{q} \frac{(m+n)!}{(q+n)!}\abs{\alpha}^{2q+n}\label{Eq: a_oneband_APP}.
		\end{align}
		\label{Eq: oneband_APP}
	\end{subequations}
	Mirroring the discussion in Sec. \ref{Sec: oneband_APP_predict}, when comparing Eqs. (\ref{Eq: oneband_APP}) and (\ref{Eq: oneband_exact_appendix}) termwise in the $m_1,m_2$ sums, we see that for all three expectation values, the highest-order term in the sum over $q$ in Eq. (\ref{Eq: oneband_APP}) corresponds to Eq. (\ref{Eq: oneband_exact_appendix}), and the lower-order terms are thus the error introduced by the APP.  
	
	We also note that the highest orders of $q$ in $\mean{\hat{a}_n^2}^{(\text{APP})}$ and $[\mean{\hat{a}_n}^{(\text{APP})}]^2$ are equal. However, the lower-order terms, which are the error introduced by the APP, are not equal. Hence, when calculating the squeezing, the exact part of the APP expectation values cancel, but the error does not, and the error therefore becomes significant, as it is relative to zero.
	
	To get a quantitative estimate of this error in the quadrature variance, we consider the terms in the sum over $q$ for the expectation values $\mean{\hat{a}_n^\dag\hat{a}_n}^{(\text{APP})}-\mean{\hat{a}_n^\dag}^{(\text{APP})}\mean{\hat{a}_n}^{(\text{APP})}$ and $\mean{\hat{a}_n^2}^{(\text{APP})}-[\mean{\hat{a}_n}^{(\text{APP})}]^2$ that are of highest-order in $\abs{\alpha}$ and are nonvanishing. This is the order $\abs{\alpha}^{2(m_1+m_2+n-1)}$. Writing these terms out using Eq. (\ref{Eq: oneband_APP}) we find
	\begin{align}
		&\mean{\hat{a}_n^\dag\hat{a}_n}^{(\text{APP})}-\mean{\hat{a}_n^\dag}^{(\text{APP})}\mean{\hat{a}_n}^{(\text{APP})} 
		\notag \\
		&\qquad\approx G_n^2 \sum_{m_1}^\infty\sum_{m_2}^\infty D_{m_1}^{(n)}D_{m_2}^{(n)} \abs{\alpha}^{2(m_1+m_2+n-1)}
		\notag \\
		&\qquad\quad\times\big(n^2+2m_1m_2+nm_2+nm_2\big)
		\label{Eq: squeez_first_term_highest_error}
	\end{align}
	and
	\begin{align}
		&\mean{\hat{a}_n^2}^{(\text{APP})}-\big[\mean{\hat{a}_n}^{(\text{APP})} \big]^2
		\notag \\
		&\qquad\approx G_n^2 e^{2in\varphi_\alpha} \sum_{m_1}^\infty\sum_{m_2}^\infty D_{m_1}^{(n)}D_{m_2}^{(n)} \abs{\alpha}^{2(m_1+m_2+n-1)}
		\notag \\
		&\qquad\quad\times\big(2m_1m_2+nm_2+nm_2\big).
		\label{Eq: squeez_second_term_highest_error}
	\end{align}
	To evaluate these, we introduce another function
	\begin{align}
		A_{n,\mu}(x)=\sum_{m=0}^\infty \frac{(-1)^m m^\mu}{m!(m+n)!}\Big(\frac{x}{2}\Big)^{2m+n},
	\end{align}
	where $x\in \mathbb{R}$, $\mu\in\mathbb{N}$, and we see that $A_{n,0}(x)=J_n(x)$. We use that
	\begin{align}
		\frac{1}{2}\Big(x\frac{d}{dx}-n\Big)\Big(\frac{x}{2}\Big)^{2m+n} = m\Big(\frac{x}{2}\Big)^{2m+n},
	\end{align}
	which means that
	\begin{align}
		A_{n,\mu}(x) = \frac{1}{2}\Big(x\frac{d}{dx}-n\Big)^\mu A_{n,0}(x)
		\notag \\
		= \frac{1}{2}\Big(x\frac{d}{dx}-n\Big)^\mu J_n(x).
	\end{align}
	From this, we get, by using identities for differentiation of the Bessel functions, that
	\begin{align}
		&A_{n,1}(x) = -\frac{x}{2}J_{n+1}(x) \\
		&A_{n,2}(x) = \frac{1}{4}\big(x^2J_{n+2}(x)-J_{n+1}(x)\big) \\
		&A_{n,3}(x) = \frac{1}{8}\big(-x^3J_{n+3}(x)+3x^2J_{n+2}(x)-J_{n+1}(x)\big).
	\end{align}
	Using this, we see that
	\begin{align}
		&G_n^2\sum_{m_1}^\infty\sum_{m_2}^\infty D_{m_1}^{(n)}D_{m_2}^{(n)}\abs{\alpha}^{2(m_1+m_2+n)}m_1^\mu m_2^\nu
		\notag \\
		&=G_n^2 \bigg[
			\sum_{l_1}C_{l_1}\sum_{m_1}^\infty \frac{(-1)^{m_1} m_1^\mu}{m_1!(m_1+n)!}\Big(\frac{\tilde{g}_0 l_1 \abs{\alpha}}{2}\Big)^{2m_1+n}
		\bigg]
		\notag\\
		&\quad\:\:\times \bigg[
		\sum_{l_2}C_{l_2}\sum_{m_2}^\infty \frac{(-1)^{m_2} m_2^\nu}{m_2!(m_2+n)!}\Big(\frac{\tilde{g}_0 l_2 \abs{\alpha}}{2}\Big)^{2m_2+n}
		\bigg]
		\notag \\
		&= G_n^2 \sum_{l_1,l_2}C_{l_1}C_{l_2}A_{n,\mu}\big(\tilde{g}_0 l_1\abs{\alpha}\big)A_{n,\nu}\big(\tilde{g}_0 l_2\abs{\alpha}\big).
		\label{Eq: expression_sum_with_ms}
	\end{align}
	Inserting Eq. (\ref{Eq: expression_sum_with_ms}) into Eqs. (\ref{Eq: squeez_first_term_highest_error}) and (\ref{Eq: squeez_second_term_highest_error}), we get the expressions
	\begin{align}
		&\mean{\hat{a}_n^\dag\hat{a}_n}^{(\text{APP})}-\mean{\hat{a}_n^\dag}^{(\text{APP})}\mean{\hat{a}_n}^{(\text{APP})} 
		\notag \\
		&\qquad\approx \frac{G_n^2}{\abs{\alpha}^2} \sum_{l_1,l_2}C_{l_1}C_{l_2} \bigg[ n^2 A_{n,0}(\tilde{g}_0 l_1\abs{\alpha})A_{n,0}(\tilde{g}_0 l_2\abs{\alpha})
		\notag \\
		&\qquad\qquad+2A_{n,1}(\tilde{g}_0 l_1\abs{\alpha})A_{n,1}(\tilde{g}_0 l_2\abs{\alpha})
		\notag\\
		&\qquad\qquad+nA_{n,1}(\tilde{g}_0 l_1\abs{\alpha})A_{n,0}(\tilde{g}_0 l_2\abs{\alpha})
		\notag\\
		&\qquad\qquad+nA_{n,0}(\tilde{g}_0 l_1\abs{\alpha})A_{n,1}(\tilde{g}_0 l_2\abs{\alpha})
		\bigg]
		\label{Eq: squeez_first_term_highest_error_analytic}
	\end{align}
	and
	\begin{align}
		&\mean{\hat{a}_n^2}^{(\text{APP})}-\big[\mean{\hat{a}_n}^{(\text{APP})} \big]^2 
		\notag \\
		&\quad\approx \frac{G_n^2}{\abs{\alpha}^2}e^{2in\varphi_\alpha} \sum_{l_1,l_2}C_{l_1}C_{l_2} \bigg[ 2A_{n,1}(\tilde{g}_0 l_1\abs{\alpha})A_{n,1}(\tilde{g}_0 l_2\abs{\alpha})
		\notag\\
		&\qquad\qquad+nA_{n,1}(\tilde{g}_0 l_1\abs{\alpha})A_{n,0}(\tilde{g}_0 l_2\abs{\alpha})
		\notag\\
		&\qquad\qquad+nA_{n,0}(\tilde{g}_0 l_1\abs{\alpha})A_{n,1}(\tilde{g}_0 l_2\abs{\alpha})
		\bigg].
		\label{Eq: squeez_second_term_highest_error_analytic}
	\end{align}
	These expressions are analytic expressions for the highest-order error in the quadrature variance, allowing for simple evaluation, which is shown in Fig. \ref{Fig: coherent_oneband_squeezing}.

	Likewise, the second-highest-order error can be computed, which is also shown in Fig. \ref{Fig: coherent_oneband_squeezing}.

	\newpage

	\bibliography{big_bibliography}

@misc{Sennary2026,
	title = {Ultrafast Quantum Optics with Attosecond Control},
	author = {Sennary, Mohamed and {Rivera-Dean}, Javier and Lewenstein, Maciej and Hassan, Mohammed Th},
	year = 2026,
	month = jan,
	number = {arXiv:2601.08671},
	eprint = {2601.08671},
	primaryclass = {physics},
	publisher = {arXiv},
	doi = {10.48550/arXiv.2601.08671},
	urldate = {2026-01-15},
	archiveprefix = {arXiv}
}

@misc{Mao2025,
	title = {Benchmarking {{Atomic Ionization Driven}} by {{Strong Quantum Light}}},
	author = {Mao, Yi-Jia and Zhou, En-Rui and Li, Yang and He, Pei-Lun and He, Feng},
	year = 2025,
	month = dec,
	number = {arXiv:2512.15458},
	eprint = {2512.15458},
	primaryclass = {quant-ph},
	publisher = {arXiv},
	doi = {10.48550/arXiv.2512.15458},
	urldate = {2026-01-15},
	archiveprefix = {arXiv}
}

@misc{Petrovic2026,
	title = {Generation of Circular Polarized High-Order Harmonics from Single Color Quantum Light},
	author = {Petrovic, Lidija and Stammer, Philipp and Lewenstein, Maciej and {Rivera-Dean}, Javier},
	year = 2026,
	month = jan,
	number = {arXiv:2601.01611},
	eprint = {2601.01611},
	primaryclass = {quant-ph},
	publisher = {arXiv},
	doi = {10.48550/arXiv.2601.01611},
	urldate = {2026-01-15},
	archiveprefix = {arXiv}
}

@misc{Lange2025c,
	title = {High-{{Order Harmonic Generation}} with {{Beyond-Semiclassical Emitter Dynamics}}: {{A Strong-Field Quantum Optical Heisenberg Picture Approach}}},
	shorttitle = {High-{{Order Harmonic Generation}} with {{Beyond-Semiclassical Emitter Dynamics}}},
	author = {Lange, Christian Saugbjerg and Lassen, Ella Elisabeth and Gothelf, Rasmus Vesterager and Madsen, Lars Bojer},
	year = 2025,
	month = dec,
	number = {arXiv:2512.14174},
	eprint = {2512.14174},
	primaryclass = {quant-ph},
	publisher = {arXiv},
	doi = {10.48550/arXiv.2512.14174},
	urldate = {2025-12-17},
	archiveprefix = {arXiv}
}

@misc{Gonzalez-Monge2025,
	title = {High-Harmonic Generation Driven by Temporal-Mode Quantum States of Light},
	author = {{Gonz{\'a}lez-Monge}, Juan M. and Feist, Johannes},
	year = 2025,
	month = dec,
	number = {arXiv:2512.06602},
	eprint = {2512.06602},
	primaryclass = {quant-ph},
	publisher = {arXiv},
	doi = {10.48550/arXiv.2512.06602},
	urldate = {2025-12-10},
	archiveprefix = {arXiv}
}

@article{Hudson1974,
	title = {When Is the {{Wigner}} Quasi-Probability Density Non-Negative?},
	author = {Hudson, R. L.},
	year = 1974,
	month = oct,
	journal = {Reports on Mathematical Physics},
	volume = {6},
	number = {2},
	pages = {249--252},
	issn = {0034-4877},
	doi = {10.1016/0034-4877(74)90007-X},
	urldate = {2025-12-08}
}

@article{Andersen2024,
  title = {Intra- and Intercycle Analysis of Intraband High-Order Harmonic Generation},
  author = {Andersen, Asbj{\o}rn Torn{\o}e and Jensen, Simon Vendelbo Bylling and Madsen, Lars Bojer},
  year = {2024},
  month = jun,
  journal = {Physical Review A},
  volume = {109},
  number = {6},
  pages = {063109},
  publisher = {American Physical Society},
  doi = {10.1103/PhysRevA.109.063109},
  urldate = {2024-10-10},
}

@article{Drummond1980,
  title = {Generalised {{P-representations}} in Quantum Optics},
  author = {Drummond, P. D. and Gardiner, C. W.},
  year = {1980},
  month = jul,
  journal = {Journal of Physics A: Mathematical and General},
  volume = {13},
  number = {7},
  pages = {2353},
  issn = {0305-4470},
  doi = {10.1088/0305-4470/13/7/018},
  urldate = {2024-10-10},
  langid = {english}
}

@article{EvenTzur2023,
  title = {Photon-Statistics Force in Ultrafast Electron Dynamics},
  author = {Even Tzur, Matan and Birk, Michael and Gorlach, Alexey and Kr{\"u}ger, Michael and Kaminer, Ido and Cohen, Oren},
  year = {2023},
  month = jun,
  journal = {Nature Photonics},
  volume = {17},
  number = {6},
  pages = {501--509},
  publisher = {Nature Publishing Group},
  issn = {1749-4893},
  doi = {10.1038/s41566-023-01209-w},
  urldate = {2024-10-10},
  copyright = {2023 The Author(s), under exclusive licence to Springer Nature Limited},
  langid = {english}
}

@book{Gerry2004,
  title = {Introductory {{Quantum Optics}}},
  author = {Gerry, Christopher and Knight, Peter},
  year = {2004},
  publisher = {Cambridge University Press},
  address = {Cambridge},
  doi = {10.1017/CBO9780511791239},
  urldate = {2024-10-10},
  isbn = {978-0-521-52735-4}
}

@article{Gorlach2020,
  title = {The Quantum-Optical Nature of High Harmonic Generation},
  author = {Gorlach, Alexey and Neufeld, Ofer and Rivera, Nicholas and Cohen, Oren and Kaminer, Ido},
  year = {2020},
  month = sep,
  journal = {Nature Communications},
  volume = {11},
  number = {1},
  pages = {4598},
  publisher = {Nature Publishing Group},
  issn = {2041-1723},
  doi = {10.1038/s41467-020-18218-w},
  urldate = {2024-10-10},
  copyright = {2020 The Author(s)},
  langid = {english}
}

@article{Gorlach2023,
  title = {High-Harmonic Generation Driven by Quantum Light},
  author = {Gorlach, Alexey and Tzur, Matan Even and Birk, Michael and Kr{\"u}ger, Michael and Rivera, Nicholas and Cohen, Oren and Kaminer, Ido},
  year = {2023},
  month = nov,
  journal = {Nature Physics},
  volume = {19},
  number = {11},
  pages = {1689--1696},
  publisher = {Nature Publishing Group},
  issn = {1745-2481},
  doi = {10.1038/s41567-023-02127-y},
  urldate = {2024-10-10},
  copyright = {2023 The Author(s), under exclusive licence to Springer Nature Limited},
  langid = {english}
}

@article{Lange2024a,
  title = {Electron-Correlation-Induced Nonclassicality of Light from High-Order Harmonic Generation},
  author = {Lange, Christian Saugbjerg and Hansen, Thomas and Madsen, Lars Bojer},
  year = {2024},
  month = mar,
  journal = {Physical Review A},
  volume = {109},
  number = {3},
  pages = {033110},
  publisher = {American Physical Society},
  doi = {10.1103/PhysRevA.109.033110},
  urldate = {2024-10-10},
}

@article{Lange2025a,
	title = {Hierarchy of approximations for describing quantum light from high-harmonic generation: A {F}ermi-{H}ubbard-model study},
	author = {Lange, Christian Saugbjerg and Madsen, Lars Bojer},
	journal = {Phys. Rev. A},
	volume = {111},
	issue = {1},
	pages = {013113},
	numpages = {13},
	year = {2025},
	month = {Jan},
	publisher = {American Physical Society},
	doi = {10.1103/PhysRevA.111.013113},
	url = {https://link.aps.org/doi/10.1103/PhysRevA.111.013113}
}

@article{Lewenstein2021,
  title = {Generation of Optical {{Schr{\"o}dinger}} Cat States in Intense Laser--Matter Interactions},
  author = {Lewenstein, M. and Ciappina, M. F. and Pisanty, E. and {Rivera-Dean}, J. and Stammer, P. and Lamprou, Th and Tzallas, P.},
  year = {2021},
  month = oct,
  journal = {Nature Physics},
  volume = {17},
  number = {10},
  pages = {1104--1108},
  publisher = {Nature Publishing Group},
  issn = {1745-2481},
  doi = {10.1038/s41567-021-01317-w},
  urldate = {2024-10-21},
  copyright = {2021 The Author(s), under exclusive licence to Springer Nature Limited},
  langid = {english}
}

@article{Manceau2019,
  title = {Indefinite-{{Mean Pareto Photon Distribution}} from {{Amplified Quantum Noise}}},
  author = {Manceau, Mathieu and Spasibko, Kirill Yu. and Leuchs, Gerd and Filip, Radim and Chekhova, Maria V.},
  year = {2019},
  month = sep,
  journal = {Physical Review Letters},
  volume = {123},
  number = {12},
  pages = {123606},
  publisher = {American Physical Society},
  doi = {10.1103/PhysRevLett.123.123606},
  urldate = {2024-10-21}
}

@article{Sharapova2020,
	title = {Properties of bright squeezed vacuum at increasing brightness},
	author = {Sharapova, P. R. and Frascella, G. and Riabinin, M. and P\'erez, A. M. and Tikhonova, O. V. and Lemieux, S. and Boyd, R. W. and Leuchs, G. and Chekhova, M. V.},
	journal = {Phys. Rev. Res.},
	volume = {2},
	issue = {1},
	pages = {013371},
	numpages = {9},
	year = {2020},
	month = {Mar},
	publisher = {American Physical Society},
	doi = {10.1103/PhysRevResearch.2.013371},
	url = {https://link.aps.org/doi/10.1103/PhysRevResearch.2.013371}
}

@article{Rasputnyi2024,
	author = {Rasputnyi, Andrei and Chen, Zhaopin and Birk, Michael and Cohen, Oren and Kaminer, Ido and Kr{\"u}ger, Michael and Seletskiy, Denis and Chekhova, Maria and Tani, Francesco},
	da = {2024/12/01},
	doi = {10.1038/s41567-024-02659-x},
	isbn = {1745-2481},
	journal = {Nature Physics},
	number = {12},
	pages = {1960--1965},
	title = {High-harmonic generation by a bright squeezed vacuum},
	ty = {JOUR},
	url = {https://doi.org/10.1038/s41567-024-02659-x},
	volume = {20},
	year = {2024}
	}

@book{Scully1997,
  title = {Quantum {{Optics}}},
  author = {Scully, Marlan O. and Zubairy, M. Suhail},
  year = {1997},
  publisher = {Cambridge University Press},
  address = {Cambridge},
  doi = {10.1017/CBO9780511813993},
  urldate = {2024-10-10},
  isbn = {978-0-521-43595-6}
}

@article{Stammer2022,
  title = {High {{Photon Number Entangled States}} and {{Coherent State Superposition}} from the {{Extreme Ultraviolet}} to the {{Far Infrared}}},
  author = {Stammer, Philipp and {Rivera-Dean}, Javier and Lamprou, Theocharis and Pisanty, Emilio and Ciappina, Marcelo F. and Tzallas, Paraskevas and Lewenstein, Maciej},
  year = {2022},
  month = mar,
  journal = {Physical Review Letters},
  volume = {128},
  number = {12},
  pages = {123603},
  publisher = {American Physical Society},
  doi = {10.1103/PhysRevLett.128.123603},
  urldate = {2024-10-10}
}

@article{Stammer2024,
  title = {Absence of Quantum Optical Coherence in High Harmonic Generation},
  author = {Stammer, Philipp},
  year = {2024},
  month = aug,
  journal = {Physical Review Research},
  volume = {6},
  number = {3},
  pages = {L032033},
  publisher = {American Physical Society},
  doi = {10.1103/PhysRevResearch.6.L032033},
  urldate = {2024-10-10}
}

@article{Stammer2024a,
  title = {Entanglement and {{Squeezing}} of the {{Optical Field Modes}} in {{High Harmonic Generation}}},
  author = {Stammer, Philipp and {Rivera-Dean}, Javier and Maxwell, Andrew S. and Lamprou, Theocharis and {Arg{\"u}ello-Luengo}, Javier and Tzallas, Paraskevas and Ciappina, Marcelo F. and Lewenstein, Maciej},
  year = {2024},
  month = apr,
  journal = {Physical Review Letters},
  volume = {132},
  number = {14},
  pages = {143603},
  publisher = {American Physical Society},
  doi = {10.1103/PhysRevLett.132.143603},
  urldate = {2024-10-10}
}

@article{Theidel2024,
	title = {Evidence of the Quantum Optical Nature of High-Harmonic Generation},
	author = {Theidel, David and Cotte, Viviane and Sondenheimer, Ren\'e and Shiriaeva, Viktoriia and Froidevaux, Marie and Severin, Vladislav and Merdji-Larue, Adam and Mosel, Philip and Fr\"ohlich, Sven and Weber, Kim-Alessandro and Morgner, Uwe and Kovacev, Milutin and Biegert, Jens and Merdji, Hamed},
	journal = {PRX Quantum},
	volume = {5},
	issue = {4},
	pages = {040319},
	numpages = {8},
	year = {2024},
	month = {Nov},
	publisher = {American Physical Society},
	doi = {10.1103/PRXQuantum.5.040319},
	url = {https://link.aps.org/doi/10.1103/PRXQuantum.5.040319}
}

@article{Vampa2015a,
  title = {Semiclassical Analysis of High Harmonic Generation in Bulk Crystals},
  author = {Vampa, G. and McDonald, C. R. and Orlando, G. and Corkum, P. B. and Brabec, T.},
  year = {2015},
  month = feb,
  journal = {Physical Review B},
  volume = {91},
  number = {6},
  pages = {064302},
  publisher = {American Physical Society},
  doi = {10.1103/PhysRevB.91.064302},
  urldate = {2024-10-10}
}

@book{Walls2008,
  title = {Quantum {{Optics}}},
  author = {Walls, D.F. and Milburn, Gerard J.},
  year = {2008},
  publisher = {Springer Berlin},
  address = {Heidelberg},
  doi = {10.1007/978-3-540-28574-8},
  urldate = {2024-10-10},
  copyright = {http://www.springer.com/tdm},
  isbn = {978-3-540-28573-1},
  langid = {english}
}

@article{Yi2024,
  title = {Generation of Massively Entangled Bright States of Light during Harmonic Generation in Resonant Media},
  author = {Yi, Sili and Klimkin, Nikolai D. and Brown, Graham Gardiner and Smirnova, Olga and Patchkovskii, Serguei and Babushkin, Ihar and Ivanov, Misha},
  journal = {Phys. Rev. X},
  volume = {15},
  issue = {1},
  pages = {011023},
  numpages = {17},
  year = {2025},
  month = {Feb},
  publisher = {American Physical Society},
  doi = {10.1103/PhysRevX.15.011023},
  url = {https://link.aps.org/doi/10.1103/PhysRevX.15.011023}
}

@article{FangLiu2023,
	title = {Strong-Field Ionization of Hydrogen Atoms with Quantum Light},
	author = {Fang, Yiqi and Sun, Feng-Xiao and He, Qiongyi and Liu, Yunquan},
	journal = {Phys. Rev. Lett.},
	volume = {130},
	issue = {25},
	pages = {253201},
	numpages = {6},
	year = {2023},
	month = {Jun},
	publisher = {American Physical Society},
	doi = {10.1103/PhysRevLett.130.253201},
	url = {https://link.aps.org/doi/10.1103/PhysRevLett.130.253201}
}

@article{EvenTzur2024,
	title = {Generation of squeezed high-order harmonics},
	author = {Tzur, Matan Even and Birk, Michael and Gorlach, Alexey and Kaminer, Ido and Kr\"uger, Michael and Cohen, Oren},
	journal = {Phys. Rev. Res.},
	volume = {6},
	issue = {3},
	pages = {033079},
	numpages = {8},
	year = {2024},
	month = {Jul},
	publisher = {American Physical Society},
	doi = {10.1103/PhysRevResearch.6.033079},
	url = {https://link.aps.org/doi/10.1103/PhysRevResearch.6.033079}
}

@article{Gothelf2025,
	title = {High-order harmonic generation in a crystal driven by quantum light},
	author = {Gothelf, Rasmus Vesterager and Lange, Christian Saugbjerg and Madsen, Lars Bojer},
	journal = {Phys. Rev. A},
	volume = {111},
	issue = {6},
	pages = {063105},
	numpages = {16},
	year = {2025},
	month = {Jun},
	publisher = {American Physical Society},
	doi = {10.1103/PhysRevA.111.063105},
}

@article{Liu2025,
	title = {Atomic Double Ionization with Quantum Light},
	author = {Liu, Haoyu and Zhang, Hanxu and Wang, Xu and Yuan, Jianmin},
	journal = {Phys. Rev. Lett.},
	volume = {134},
	issue = {12},
	pages = {123202},
	numpages = {7},
	year = {2025},
	month = {Mar},
	publisher = {American Physical Society},
	doi = {10.1103/PhysRevLett.134.123202},
}

@article{Wang2023,
	title = {High-order above-threshold ionization of an atom in intense quantum light},
	author = {Wang, ShiJun and Lai, XuanYang},
	journal = {Phys. Rev. A},
	volume = {108},
	issue = {6},
	pages = {063101},
	numpages = {6},
	year = {2023},
	month = {Dec},
	publisher = {American Physical Society},
	doi = {10.1103/PhysRevA.108.063101},
	url = {https://link.aps.org/doi/10.1103/PhysRevA.108.063101}
}

@article{Heimerl2024,
	author = {Heimerl, Jonas and Mikhaylov, Alexander and Meier, Stefan and H{\"o}llerer, Henrick and Kaminer, Ido and Chekhova, Maria and Hommelhoff, Peter},
	doi = {10.1038/s41567-024-02472-6},
	isbn = {1745-2481},
	journal = {Nature Physics},
	number = {6},
	pages = {945--950},
	title = {Multiphoton electron emission with non-classical light},
	ty = {JOUR},
	volume = {20},
	year = {2024}}

@article{Heimerl2025,
	author = {Heimerl, Jonas and Rasputnyi, Andrei and P{\"o}lloth, Jonathan and Meier, Stefan and Chekhova, Maria and Hommelhoff, Peter},
	da = {2025/11/07},
	id = {Heimerl2025},
	isbn = {1745-2481},
	journal = {Nature Physics},
	title = {Quantum light drives electrons strongly at metal needle tips},
	ty = {JOUR},
	year = {2025},
	}

@article{Lange2025b,
	title = {Excitonic Enhancement of Squeezed Light in Quantum-Optical High-Harmonic Generation from a {M}ott Insulator},
	author = {Lange, Christian Saugbjerg and Hansen, Thomas and Madsen, Lars Bojer},
	journal = {Phys. Rev. Lett.},
	volume = {135},
	issue = {4},
	pages = {043603},
	numpages = {7},
	year = {2025},
	month = {Jul},
	publisher = {American Physical Society},
	doi = {10.1103/wyk5-k8tk},
}

@misc{Stammer2025,
	title={Theory of quantum optics and optical coherence in high harmonic generation}, 
	author={Philipp Stammer and Javier Rivera-Dean and Maciej Lewenstein},
	year={2025},
	eprint={2504.13287},
	archivePrefix={arXiv},
	primaryClass={quant-ph},
}

@article{Chekhova2015,
	title = {Bright squeezed vacuum: Entanglement of macroscopic light beams},
	journal = {Optics Communications},
	volume = {337},
	pages = {27-43},
	year = {2015},
	note = {Macroscopic quantumness: theory and applications in optical sciences},
	issn = {0030-4018},
	doi = {https://doi.org/10.1016/j.optcom.2014.07.050},
	url = {https://www.sciencedirect.com/science/article/pii/S0030401814006695},
	author = {M.V. Chekhova and G. Leuchs and M. Żukowski},
}

@article{Agafonov2010,
	title = {Two-color bright squeezed vacuum},
	author = {Agafonov, Ivan N. and Chekhova, Maria V. and Leuchs, Gerd},
	journal = {Phys. Rev. A},
	volume = {82},
	issue = {1},
	pages = {011801},
	numpages = {4},
	year = {2010},
	month = {Jul},
	publisher = {American Physical Society},
	doi = {10.1103/PhysRevA.82.011801},
	url = {https://link.aps.org/doi/10.1103/PhysRevA.82.011801}
}

@article{Perez2014,
	author = {A. M. P\'{e}rez and T. Sh. Iskhakov and P. Sharapova and S. Lemieux and O. V. Tikhonova and M. V. Chekhova and G. Leuchs},
	journal = {Opt. Lett.},
	number = {8},
	pages = {2403--2406},
	publisher = {Optica Publishing Group},
	title = {Bright squeezed-vacuum source with 1.1 spatial mode},
	volume = {39},
	month = {Apr},
	year = {2014},
	url = {https://opg.optica.org/ol/abstract.cfm?URI=ol-39-8-2403},
	doi = {10.1364/OL.39.002403},
}

@article{Finger2015,
	title = {Raman-Free, Noble-Gas-Filled Photonic-Crystal Fiber Source for Ultrafast, Very Bright Twin-Beam Squeezed Vacuum},
	author = {Finger, Martin A. and Iskhakov, Timur Sh. and Joly, Nicolas Y. and Chekhova, Maria V. and Russell, Philip St. J.},
	journal = {Phys. Rev. Lett.},
	volume = {115},
	issue = {14},
	pages = {143602},
	numpages = {5},
	year = {2015},
	month = {Sep},
	publisher = {American Physical Society},
	doi = {10.1103/PhysRevLett.115.143602},
	url = {https://link.aps.org/doi/10.1103/PhysRevLett.115.143602}
}

@article{Iskhakov2012,
	author = {T. Sh. Iskhakov and A. M. P\'{e}rez and K. Yu. Spasibko and M. V. Chekhova and G. Leuchs},
	journal = {Opt. Lett.},
	number = {11},
	pages = {1919--1921},
	publisher = {Optica Publishing Group},
	title = {Superbunched bright squeezed vacuum state},
	volume = {37},
	month = {Jun},
	year = {2012},
	url = {https://opg.optica.org/ol/abstract.cfm?URI=ol-37-11-1919},
	doi = {10.1364/OL.37.001919},
}

@article{Wang2025,
	title = {High-order harmonic generation in quantum light by a generalized von {{Neumann}} lattice method},
	author = {Wang, Yi-Ben and Bian, Xue-Bin},
	journal = {Phys. Rev. A},
	volume = {111},
	issue = {4},
	pages = {043111},
	numpages = {14},
	year = {2025},
	month = {Apr},
	publisher = {American Physical Society},
	doi = {10.1103/PhysRevA.111.043111},
	url = {https://link.aps.org/doi/10.1103/PhysRevA.111.043111}
}

@misc{Rivera-Dean2025a,
	title={Attosecond quantum optical interferometry}, 
	author={Javier Rivera-Dean and Lidija Petrovic and Maciej Lewenstein and Philipp Stammer},
	year={2025},
	eprint={2511.01097},
	archivePrefix={arXiv},
	primaryClass={quant-ph},
	url={https://arxiv.org/abs/2511.01097}, 
}

@article{Lyu2025,
	title = {Effect of photon quantum statistics on electrons in above-threshold ionization},
	author = {Lyu, Zijian and Sun, Fengxiao and Fang, Yiqi and He, Qiongyi and Liu, Yunquan},
	journal = {Phys. Rev. Res.},
	volume = {7},
	issue = {1},
	pages = {L012072},
	numpages = {7},
	year = {2025},
	month = {Mar},
	publisher = {American Physical Society},
	doi = {10.1103/PhysRevResearch.7.L012072},
	url = {https://link.aps.org/doi/10.1103/PhysRevResearch.7.L012072}
}

@misc{Stammer2025b,
	title={Weak measurement in strong laser field physics}, 
	author={Philipp Stammer and Javier Rivera-Dean and Marcelo F. Ciappina and Maciej Lewenstein},
	year={2025},
	eprint={2508.09048},
	archivePrefix={arXiv},
	primaryClass={quant-ph},
	url={https://arxiv.org/abs/2508.09048}, 
}

@misc{Rivera-Dean2025c,
	title={Microscopic analysis of above-threshold ionization driven by squeezed light}, 
	author={J. Rivera-Dean and P. Stammer and C. Figueira de Morisson Faria and M. Lewenstein},
	year={2025},
	eprint={2508.01621},
	archivePrefix={arXiv},
	primaryClass={quant-ph},
	url={https://arxiv.org/abs/2508.01621}, 
}
	
\end{document}